\documentclass[review]{elsarticle}
\usepackage{geometry}
\usepackage{enumitem}
\usepackage{lineno,hyperref}
\modulolinenumbers[5]

\usepackage{booktabs}
\usepackage{longtable}
\usepackage{xcolor}
\usepackage{array,multirow,graphicx}
\usepackage{float}
\usepackage{makecell}
\usepackage[T1]{fontenc}
\usepackage[utf8]{inputenc}

\journal{European Journal of Operational Research}

\biboptions{authoryear}
\usepackage{amsmath}
\usepackage{amsfonts}
\usepackage{bm}
\usepackage{caption}
\usepackage{subcaption}

\usepackage{tikz}
\usepackage{forest}

\usetikzlibrary{arrows,shapes,positioning,shadows,trees}

\tikzset{
  basic/.style  = {draw, text width=2cm, drop shadow, font=\sffamily, rectangle},
  root/.style   = {basic, rounded corners=2pt, thin, align=center,
                   fill=green!30},
  level 2/.style = {basic, rounded corners=6pt, thin,align=center, fill=green!60,
                   text width=4em},
  level 3/.style = {basic, thin, align=left, fill=pink!60, text width=1.5em}
}
\newcommand{\relation}[3]
{
	\draw (#3.south) -- +(0,-#1) -| ($ (#2.north) $)
}

\makeatletter
\def\widebreve{\mathpalette\wide@breve}
\def\wide@breve#1#2{\sbox\z@{$#1#2$}%
	\mathop{\vbox{\m@th\ialign{##\crcr
				\kern0.08em\brevefill#1{0.8\wd\z@}\crcr\noalign{\nointerlineskip}%
				$\hss#1#2\hss$\crcr}}}\limits}
\def\brevefill#1#2{$\m@th\sbox\tw@{$#1($}%
	\hss\resizebox{#2}{\wd\tw@}{\rotatebox[origin=c]{90}{\upshape(}}\hss$}
\makeatother

\makeatletter
\DeclareRobustCommand\widecheck[1]{{\mathpalette\@widecheck{#1}}}
\def\@widecheck#1#2{%
	\setbox\z@\hbox{\m@th$#1#2$}%
	\setbox\tw@\hbox{\m@th$#1%
		\widehat{%
			\vrule\@width\z@\@height\ht\z@
			\vrule\@height\z@\@width\wd\z@}$}%
	\dp\tw@-\ht\z@
	\@tempdima\ht\z@ \advance\@tempdima2\ht\tw@ \divide\@tempdima\thr@@
	\setbox\tw@\hbox{%
		\raise\@tempdima\hbox{\scalebox{1}[-1]{\lower\@tempdima\box
				\tw@}}}%
	{\ooalign{\box\tw@ \cr \box\z@}}}
\makeatother

\usepackage{array}
\usepackage{scalerel,stackengine}
\stackMath
\newcommand\reallywidehat[1]{%
\savestack{\tmpbox}{\stretchto{%
  \scaleto{%
    \scalerel*[\widthof{\ensuremath{#1}}]{\kern-.6pt\bigwedge\kern-.6pt}%
    {\rule[-\textheight/2]{1ex}{\textheight}}
  }{\textheight}%
}{0.5ex}}%
\stackon[1pt]{#1}{\tmpbox}%
}

\newcommand{\lambdavet}{\mathbf{\lambda}}
\newcommand{\avet}{\textbf{a}}
\newcommand{\bvet}{\textbf{b}}

\newcommand{\pvet}{\textbf{p}}

\newcommand{\yvet}{\textbf{y}}
\newcommand{\Avet}{\textbf{A}}

\newcommand{\Cvet}{\textbf{C}}

\newcommand{\Gvet}{\textbf{G}}

\newcommand{\Ivet}{\textbf{I}}
\newcommand{\Jvet}{\textbf{J}}

\newcommand{\Lvet}{\textbf{L}}
\newcommand{\Mvet}{\textbf{M}}

\newcommand{\Pvet}{\textbf{P}}

\newcommand{\Svet}{\textbf{S}}

\newcommand{\Uvet}{\textbf{U}}
\newcommand{\Wvet}{\textbf{W}}

\newcommand{\Zerovet}{\textbf{0}}
\bmdefine{\xhvet}{\mathsf{x}}
\bmdefine{\yhvet}{\mathsf{y}}
\bmdefine{\Chvet}{\mathsf{C}}
\bmdefine{\Ghvet}{\mathsf{G}}
\bmdefine{\Jhvet}{\mathsf{J}}
\bmdefine{\Shvet}{\mathsf{S}}
\bmdefine{\Uhvet}{\mathsf{U}}
\bmdefine{\Gammavet}{\mathsf{\Gamma}}
\bmdefine{\Omegavet}{\mathsf{\Omega}}
\bmdefine{\Thetavet}{\mathsf{\Theta}}
\bmdefine{\ahvet}{\mathfrak{a}}
\bmdefine{\bhvet}{\mathfrak{b}}

\DeclareMathOperator*{\argmin}{arg\,min}
\begin{document}

\begin{frontmatter}

\title{Forecast combination based forecast reconciliation: insights and extensions}



\author[mymainaddress]{Tommaso Di Fonzo\corref{mycorrespondingauthor}}
\cortext[mycorrespondingauthor]{Corresponding author}
\ead{difonzo@stat.unipd.it}

\author[mymainaddress]{Daniele Girolimetto}

\address[mymainaddress]{Department of Statistical Sciences, University of Padua, \\ Via C. Battisti 241, 35121 Padova (Italy)}

\begin{abstract}
In a recent paper, while elucidating the links between forecast combination and cross-sectional forecast reconciliation, \cite{Hollyman2021} have proposed a forecast combination-based approach to the reconciliation of a simple hierarchy. A new Level Conditional Coherent ($LCC$) point forecast reconciliation procedure was developed, and it was shown that the simple average of a set of $LCC$, and bottom-up reconciled forecasts (called Combined Conditional Coherent, $CCC$) results in good performance as compared to those obtained through the state-of-the-art cross-sectional reconciliation procedures. 
In this paper, we build upon and extend 
this proposal
along some new directions.
(1) We shed light on the nature and the mathematical derivation of the $LCC$ reconciliation formula, showing that it is the result of an exogenously linearly constrained  minimization of a quadratic loss function in the differences between the target and the base forecasts with a diagonal associated matrix. (2) Endogenous constraints may be considered as well, resulting in level conditional reconciled forecasts of all the involved series, where both the upper and the bottom time series are coherently revised.
We show that even in this framework it is still valid the interesting interpretation given by \cite{Hollyman2021} of the reconciliation formula as the combination of direct (base) and indirect forecasts, the latter ones depending on the accounting relationships linking upper and bottom series.
(3) As the $LCC$ procedure (i.e., with exogenous constraints, but the result holds in the endogenous case as well) does not guarantee the non-negativity of the reconciled forecasts, we argue that - when non-negativity is a natural attribute of the variables to be forecast - its interpretation as an `unbiased top-down reconciliation procedure' leaves room for some doubts. 
(4) The new procedures 
are used in a forecasting experiment on the classical Australian Tourism Demand (Visitor Nights) dataset.
Due to the crucial role played by the (possibly different) models used to compute the base forecasts, we re-interpret the $CCC$ reconciliation of \cite{Hollyman2021} as a forecast pooling approach, showing that accuracy improvement may be gained by adopting a simple forecast averaging strategy.
\end{abstract}


\end{frontmatter}

\newpage

\section{Introduction}
\label{sec:1}
A hierarchical/grouped time series is a linearly constrained multiple time series consisting of a collection of time series that follows a hierarchical aggregation structure (\citealp{Panagiotelis2020a}).
As an example, sales data can be disaggregated by product categories, and then by product subcategories, down to Stock Keeping Unit (SKU). In order to provide the appropriate demand forecast information given various managerial levels and functional disciplines within organizations, reliance on hierarchical forecasting is increasing, with the objective of producing coherent forecasts while improving their accuracy and reducing the overall forecasting burden (\citealp{Fliedner2001}).

More generally, hierarchical/grouping forecasting is the process of generating coherent forecasts (or reconciling incoherent forecasts), allowing time series to be forecast individually, but preserving the relationships within the hierarchy/group (\citealp{Hyndman2021book}, ch. 11).
It can be seen as a statistical device that can improve forecast accuracy through the use of forecast combination, whose main motivation is to provide coherent predictions for decisions at different levels of the hierarchy.

Classical reconciliation approaches are bottom-up, top-down, and middle-out (\citealp{Athanasopoulos2009}).
Bottom-up forecasting (\citealp{Dunn1976}) involves forecasting the most granular level of the hierarchy, then aggregating up to create estimates for the higher levels. The main advantage of this method is that, because forecasts are obtained at the lowest level of the hierarchy, no information is lost due to aggregation. However, it ignores the relationships between the series, and usually performs poorly on highly aggregated data. Furthermore, information at lower levels of the hierarchy tends to be noiser, potentially resulting in a reduced overall forecast accuracy.
In the top-down approach (\citealp{Gross1990}), the top level of the hierarchy is first forecast, and then this forecast is split up to get estimates for the lower levels, typically using historical or forecasted proportions (\citealp{Athanasopoulos2009}).
The middle-out approach is a combination of the bottom-up and top-down approaches: the middle level (neither the most granular nor the most aggregated) variables are forecast. These values are then used to compute the higher levels' forecasts using the bottom-up approach, and the lower levels with the top-down approach.

Modern least squares-based reconciliation techniques emerged in the cross-sectional framework (optimal combination approach, \citealp{Hyndman2011}), and have then been extended both in temporal (\citealp{Athanasopoulos2017}, \citealp{Nystrup2020}), and cross-temporal (\citealp{Kourentzes2019}, \citealp{DiFonzoGiro2020}) frameworks, designed to align short- and long-term forecasts for consistency of different planning and budgeting purposes. 
This class of techniques is usually associated to a forecasting scheme where all the time series are independently forecast at all levels (cross-sectional and/or temporal), producing incoherent base forecasts, that are then transformed using all the information and relationships a hierarchy can offer in order to be coherent along the chosen dimensions (cross-sectional, temporal, or both). This result is obtained using a linear regression model, and the newly coherent forecasts are a weighted sum of the forecasts from all levels, with the weights found by solving a system of equations ensuring the natural relationships between the different levels of the hierarchy are satisfied.
If the base forecasts are unbiased, \cite{Hyndman2011} show that the optimal combination approach provides unbiased reconciled forecasts at all levels with minimal loss of information, taking advantage of the relationships between time series to find patterns. 

In practice, the best choice is oftentimes a combination of hierarchies, as forecast accuracy tends to improve if the model can learn from multiple relationships (\citealp{Bates1969}, \citealp{Timmermann2006}). Moving from this observation, 
in a recent paper, while elucidating the links between forecast combination and cross-sectional forecast reconciliation, \cite{Hollyman2021} have proposed a forecast combination-based approach to the reconciliation of a simple hierarchy. A new Level Conditional Coherent ($LCC$) point forecast reconciliation procedure was developed, and it was shown that the simple average of a set of $LCC$, and bottom-up reconciled forecasts (called Combined Conditional Coherent, $CCC$) results in good performance as compared to those obtained through the state-of-the-art cross-sectional reconciliation procedures (\citealp{Wickramasuriya2019}). 
In this paper, we build upon and extend this proposal along some new directions.
\begin{enumerate}
\item We shed light on the nature and the mathematical derivation of the $LCC$ reconciliation formula, showing that it is the result of an exogenously linearly constrained  minimization of a quadratic loss function in the differences between the target and the base forecasts with a diagonal associated matrix.
\item Endogenous constraints may be considered as well, resulting in level conditional reconciled forecasts of all the involved series, where both the upper and the bottom time series are coherently revised.
We show that even in this framework it is still valid the interesting interpretation given by \cite{Hollyman2021} of the reconciliation formula as the combination of direct (base) and indirect forecasts, the latter ones depending on the accounting relationships linking upper and bottom series.
The extension to the cases where a full metric matrix is considered in the definition of the loss function is straightforward, which might be useful when (if) suitable error forecast covariance matrices may be estimated.
\item We show that the $LCC$ approach (i.e., with exogenous constraints, but the result holds in the endogenous case as well) does not guarantee the non-negativity of the reconciled forecasts. Thus we argue that - when the variables to be forecast are intrinsically non-negative - interpreting $LCC$ as an `unbiased top-down reconciliation procedure' (\citealp{Hollyman2021}) leaves room for some doubts. 
\item The new procedures, available in the \texttt{R} package \texttt{FoReco} (\citealp{FoReco2021}), 
are used in a forecasting experiment on the classical Australian Tourism Demand (Visitor Nights) dataset, where the original results found by \cite{Hollyman2021} are re-assessed (i) using the relative accuracy indices for multiple comparisons recommended by \cite{Davydenko2013}, and (ii) taking into account the non-negativity issues posed by the dataset at hand.
\item Finally, due to the crucial role played by the (possibly different) models used to compute the base forecasts, the $CCC$ reconciliation strategy proposed by \cite{Hollyman2021} (called $CCC_H$ in our paper) is interpreted as a forecast pooling strategy (\citealp{Hendry2004}, \citealp{Marcellino2004}, \citealp{Geweke2011}, \citealp{KourentzesBarrowPetropoulos2019}), 
and we show that the intuition behind the $CCC_H$ strategy can be further improved by adopting a somehow less arbitrary simple forecast averaging strategy.
\end{enumerate}

The paper is organized as follows. In section \ref{sec:notation} we set the notation, define the forecast reconciliation problem, and show the Level Conditional Coherent forecast reconciliation approach proposed by \cite{Hollyman2021}.
We re-interpret this procedure in terms of an optimization problem, with either exogenous (section \ref{sec:LCCexo}), or endogenous (section \ref{sec:LCCendo}) constraints, which encompasses the forecast reconciliation procedure so far as a particular case.
An empirical application is performed in section \ref{sec:VN525}, where the rolling forecast experiment performed by \cite{Hollyman2021} is reprised and extended in light of the new insights.
Section \ref{sec: conclusions} contains conclusions and indications for future research. An on-line appendix contains supplementary tables and graphs related to the empirical application. 
 
\section{Problem definition and notation}
\label{sec:notation}
Let us consider a linearly constrained multiple time series with a genuine hierarchical/grouped structure consisting of $L>0$ levels above the bottom level. When $L=1$ we face an \textit{elementary hierarchy}, formed by one top-level series and $n_b$ component bottom time series (bts).
In general, denote $n_a$ the number of the upper time series (uts) in a hierarchy, and $n_l$ the number of knots (series) of the generic level $l$ of the series, $l=1,\ldots,L$, with  $n_1=1$, and $\displaystyle\sum_{l=1}^{L}n_l = n_a$.

Consider now the $L$ hierarchical structures simply formed by the $n_b$ bts and the $n_l$ upper time series (uts) of level $l$, $l=1,\ldots, L$.
For example, the hierarchical series in the left panel of figure 1 (see appendix 2 in \citealp{Hollyman2021}) consists of three levels: the total series $T$ at level 1, series $X$ and $Y$ at the intermediate level 2, and the five bottom time series $A, B, C, D, E$, at the bottom level 3,  with $T = X+Y = A+B+C+D+E$, $X=A+B$, and $Y=C+D+E$.

\begin{figure}[ht]
	\centering
	\resizebox{0.4\linewidth}{!}{
		\begin{tikzpicture}[baseline=(current  bounding  box.center),
			every node/.append style={shape=circle,
				draw=black},
			minimum width=1cm,
			minimum height=1cm]
			
			\node at (0, 0) (A){$A$};
			\node at (1.5, 0) (B){$B$};
			\node at (3, 0) (C){$C$};
			\node at (4.5, 0) (D){$D$};
			\node at (6, 0) (E){$E$};
			\node at (0.75, 2) (X){$X$};
			\node at (4.5, 2) (Y){$Y$};
			\node at (2.75, 4) (T){$T$};
			
			\relation{0.3}{A}{X};
			\relation{0.3}{B}{X};
			\relation{0.3}{C}{Y};
			\relation{0.3}{D}{Y};
			\relation{0.3}{E}{Y};
			\relation{0.3}{X}{T};
			\relation{0.3}{Y}{T};
		\end{tikzpicture}
	} \hspace{1.5cm}
	\resizebox{0.35\linewidth}{!}{
		\begin{tikzpicture}[baseline=(current  bounding  box.center),
			every node/.append style={shape=circle,
				draw=black},
			minimum width=1cm,
			minimum height=1cm]
			
			\node at (0, 0) (A){$A$};
			\node at (1.5, 0) (B){$B$};
			\node at (3, 0) (C){$C$};
			\node at (4.5, 0) (D){$D$};
			\node at (6, 0) (E){$E$};
			\node at (0.75, 1.7) (X){$X$};
			\node at (4.5, 1.7) (Y){$Y$};
			
			\node at (0, 3) (A1){$A$};
			\node at (1.5, 3) (B1){$B$};
			\node at (3, 3) (C1){$C$};
			\node at (4.5, 3) (D1){$D$};
			\node at (6, 3) (E1){$E$};
			\node at (3, 4.7) (T){$T$};
			
			\relation{0.2}{A}{X};
			\relation{0.2}{B}{X};
			\relation{0.2}{C}{Y};
			\relation{0.2}{D}{Y};
			\relation{0.2}{E}{Y};
			
			\relation{0.2}{A1}{T};
			\relation{0.2}{B1}{T};
			\relation{0.2}{C1}{T};
			\relation{0.2}{D1}{T};
			\relation{0.2}{E1}{T};
			
		\end{tikzpicture}
	}
	\caption{A three-level hierarchy (left), and its elementary hierarchies (right)}
	\label{Fig:hts1}
\end{figure}
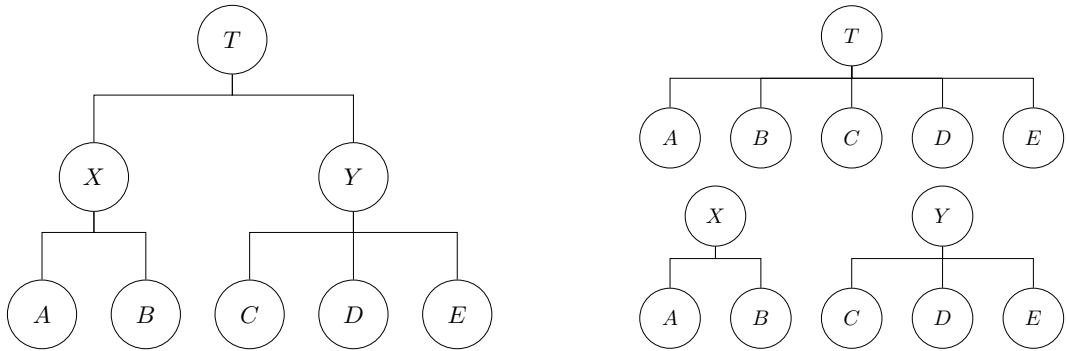

\noindent Such a series may be represented in structural form as $\yvet = \Svet\bvet$, where
$ \Svet = \begin{bmatrix}
	1 & 1 & 1 & 1 & 1 \\
	1 & 1 & 0 & 0 & 0 \\
	0 & 0 & 1 & 1 & 1 \\
	\multicolumn{5}{c}{\Ivet_5}
\end{bmatrix}$
is a $(8 \times 5)$ cross-sectional summing matrix,
$\yvet = \begin{bmatrix}
	T \; X \; Y \; A \; B \; C \; D \; E
\end{bmatrix}'$,
and
$\bvet = \begin{bmatrix}
	A \; B \; C \; D \; E
\end{bmatrix}'$.

The main series $\yvet$ contains other two linearly constrained multiple time series, sharing the same bts: $\yvet_1 = \Svet_1\bvet$ and $\yvet_2 = \Svet_2\bvet$,
with, respectively,
$$
\Svet_1 = \begin{bmatrix}
	1 & 1 & 1 & 1 & 1 \\
	\multicolumn{5}{c}{\Ivet_5}
\end{bmatrix},
\quad
\yvet_1 = \begin{bmatrix}
	T \; A \; B \; C \; D \; E
\end{bmatrix}',
$$
$$
\Svet_2 = \begin{bmatrix}
	1 & 1 & 0 & 0 & 0 \\
	0 & 0 & 1 & 1 & 1 \\
	\multicolumn{5}{c}{\Ivet_5}
\end{bmatrix},
\quad
\yvet_2 = \begin{bmatrix}
	X \; Y \; A \; B \; C \; D \; E
\end{bmatrix}'.
$$
Notice that the matrix $\Svet_1$ describes the elementary hierarchy formed by the top-level and the bottom series, not considering the intermediate level $l=2$, whereas $\Svet_2$ does not correspond to a \textit{summing matrix} of a standard structural representation of a hierarchical/grouped series, because a unique top level series is not present. In this latter case, we may recognize two distinct elementary hierarchies (right panel of figure 1): the former valid for series $X$, re-interpreted as a `top level series', and the latter for series $Y$ as well, respectively given by:
\[
\begin{bmatrix}
	X \\ A \\ B
\end{bmatrix} =
\begin{bmatrix}
	1 & 1 \\ \multicolumn{2}{c}{\Ivet_2}
\end{bmatrix}
\begin{bmatrix}
	A \\ B
\end{bmatrix},
\quad
\begin{bmatrix}
	Y \\ C \\ D \\ E
\end{bmatrix} =
\begin{bmatrix}
	1 & 1 & 1 \\ \multicolumn{3}{c}{\Ivet_3}
\end{bmatrix}
\begin{bmatrix}
	C \\ D \\ E
\end{bmatrix} .
\]

\noindent Let $\Cvet$ be the $(3 \times 5)$ cross-sectional (contemporaneous) aggregation matrix mapping the five bts into the three uts of the multiple series, which is linked to the summing matrix $\Svet$ by the relationship
$
\Svet = \begin{bmatrix}
	\Cvet' \; \; \; \Ivet_{5}
\end{bmatrix}'
$.
Matrix $\Cvet$ consists of two sub-matrices $\Cvet_l$, $l=1,2$:
%
\[
\Cvet = \begin{bmatrix}
	1 & 1 & 1 & 1 & 1 \\ 
	1 & 1 & 0 & 0 & 0 \\
	0 & 0 & 1 & 1 & 1
\end{bmatrix}
\quad
\Cvet_1 = \begin{bmatrix}
	1 & 1 & 1 & 1 & 1
\end{bmatrix}'
\quad
\Cvet_2 = \begin{bmatrix}
	1 & 1 & 0 & 0 & 0 \\
	0 & 0 & 1 & 1 & 1
\end{bmatrix} .
\]
The level-$l$ constrained multiple time series may be thus represented as $\yvet_l =\Svet_l\bvet$, $l=1,2$, where the vector $\yvet_l = \begin{bmatrix}
	\avet_l \\ \bvet
\end{bmatrix}$ has dimension $\left[(n_l+n_b) \times 1\right]$, $n_l$ being the number of knots (series) in the $\left(n_l \times 1\right)$ vector of the level-$l$ time series $\avet_l$ (i.e., $n_1=1$, and $n_2=2$), and
$
\Svet_l = \begin{bmatrix}
	\Cvet_l' \; \; \; \Ivet_{n_b}
\end{bmatrix}'$, $l=1,2$,
has dimension $\left[\left(n_l+5\right) \times 5\right]$.

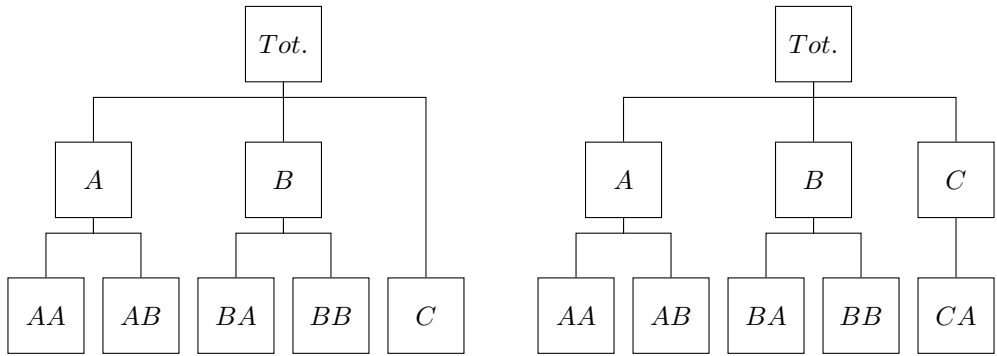
\begin{figure}[ht]
	\centering
	\begin{tabular}{ccc}
		\begin{tikzpicture} [baseline=(current  bounding  box.center),
			every node/.append style={shape=rectangle,
				draw=black,
				minimum size=1cm}
			]
			\node at (-5, 0) (AA){$AA$};
			\node at (-3.75, 0) (AB){$AB$};
			\node at (-2.5, 0) (BA){$BA$};
			\node at (-1.25, 0) (BB){$BB$};
			\node at (0, 0) (CA){$C$};
			\node at (-4.375, 1.8) (A){$A$};
			\node at (-1.875, 1.8) (B){$B$};
			\node at (-1.875, 3.6) (T){$Tot.$};
			\relation{0.2}{B}{T};
			\relation{0.2}{A}{T};
			\relation{0.2}{CA}{T};
			\relation{0.2}{BA}{B};
			\relation{0.2}{BB}{B};
			\relation{0.2}{AA}{A};
			\relation{0.2}{AB}{A};
		\end{tikzpicture} & ~ &
		\begin{tikzpicture} [baseline=(current  bounding  box.center),
			every node/.append style={shape=rectangle,
				draw=black,
				minimum size=1cm}
			]
			\node at (-5, 0) (AA){$AA$};
			\node at (-3.75, 0) (AB){$AB$};
			\node at (-2.5, 0) (BA){$BA$};
			\node at (-1.25, 0) (BB){$BB$};
			\node at (0, 0) (CA){$CA$};
			\node at (-4.375, 1.8) (A){$A$};
			\node at (-1.875, 1.8) (B){$B$};
			\node at (0, 1.8) (C){$C$};
			\node at (-1.875, 3.6) (T){$Tot.$};
			\relation{0.2}{C}{T};
			\relation{0.2}{B}{T};
			\relation{0.2}{A}{T};
			\relation{0.2}{CA}{C};
			\relation{0.2}{BA}{B};
			\relation{0.2}{BB}{B};
			\relation{0.2}{AA}{A};
			\relation{0.2}{AB}{A};
		\end{tikzpicture}
	\end{tabular}
	\caption{A simple unbalanced hierarchy (left) and its balanced version (right)}
	\label{Fig:unbal}
\end{figure}

It should be added that, in order to develop the results that follow, the hierarchy has to be \textbf{balanced}, that is, each `knot' (series) at an upper level wrt the bottom one, must have at least a `child' series.
A simple unbalanced three-level hierarchy is shown in the left panel of figure \ref{Fig:unbal}, where variable $C$ 
has no `children', and thus is considered as a bottom variable, at level three of the hierarchy.
\noindent The right panel shows the `balanced version' of the same hierarchy, where $CA = C$, and thus variable $C$ is (duplicated and)
present at both levels two and three.
 Possible duplication of some variables should be conveniently accounted for, e.g. when evaluating the reconciled forecasts' accuracy for all the series in the hierarchy. 

\subsection{Notation}
Let $l$, $l=1,\ldots,L$, the index associated to a generic level - above the bts level - of the hierarchy/grouping which characterizes the multiple time series whose base forecasts are wished to be reconciled, and assume that $l=1$ denote the top-level, consisting of the total of the whole hierarchical/grouped system.
Let $\yvet = \begin{bmatrix}
	\avet \\ \bvet
\end{bmatrix}$ be the $(n \times 1)$ vector of target forecasts, formed by the $(n_a \times 1)$ vector $\avet$ of upper time series (uts), and by the $(n_b \times 1)$ vector $\bvet$ of bottom time series (bts). Denote the base forecasts vector $\widehat{\yvet} = \begin{bmatrix}
	\widehat{\avet} \\ \widehat{\bvet}
\end{bmatrix}$. In addition, decompose vector $\widehat{\avet}$ into the sub-vectors forming each of the upper $L$ levels of the hierarchy/grouping:
$$
\widehat{\avet} = \begin{bmatrix}
	\widehat{a}_1 \\ \widehat{\avet}_2 \\ \vdots \\ \widehat{\avet}_L
\end{bmatrix} ,
$$
where $\widehat{\avet}_l$, $l=1,\ldots, L$, has dimension $(n_l \times 1)$, with $n_1=1$ and $\displaystyle\sum_{l=1}^{L} n_l = n_a$.

Denote $\Cvet_l$ the $(n_l \times n_b)$ matrix mapping the bts into the level-$l$ uts (i.e., $\avet_l = \Cvet_l\bvet$). The complete aggregation matrix $\Cvet$, mapping all the bts into the uts of all levels $l=1,\ldots,L$, may be written as
\begin{equation}
	\label{Cmat}
\Cvet = \begin{bmatrix}
	\Cvet_1 \\ \Cvet_2 \\ \vdots \\ \Cvet_L
\end{bmatrix},
\end{equation}
where the generic matrix $\Cvet_l$ is ($n_l \times n_b$), $l=1, \ldots, L$, and $\Cvet_1 = {\bf 1}_{n_b}'$ is a $(1 \times n_b)$ row vector of ones (sum vector).

\subsection{Level Conditional Coherent ($LCC$) forecast reconciliation}
\label{subsec:LCC}
The central core of the proposal by 
\cite{Hollyman2021} is a \textit{level conditional coherent} forecast reconciliation procedure which, for any level $l$ of the hierarchy, transforms the vector $\widehat{\bvet}$ of unbiased base forecasts of the bts in unbiased reconciled forecasts $\widetilde{\bvet}^{(l)}$ coherent with the \textbf{base forecasts} of the series at that specific level of the hierarchy. The exponent $^{(l)}$ highlights that the base forecasts of the bts are transformed in such a way that they are coherent with the base forecasts of the series at level $l$ (or, equivalently, that the reconciled forecasts are coherent conditional to the base forecasts of the series at level $l$).
A set of $L$ reconciled forecasts, each conditional to the $n_l$ base forecasts of a specific level-$l$, may thus be computed.
When $l=1$, i.e. the only upper level series is the total aggregate at the top of the hierarchy, \cite{Hollyman2021} interpret this procedure as a \textit{top down forecast reconciliation}, as the conditioning top-level forecast remains unchanged, just as it happens with the classical top-down reconciliation (\citealp{Gross1990}, \citealp{Athanasopoulos2009}). We will come back on this point later on.
In the remaining cases ($1 < l \le L$), in order to coherently adjust the whole vector of forecasts, the level conditional reconciled forecasts are transformed through a \textit{middle-out} reconciliation procedure, by simply pre-multiplyng the bts reconciled forecasts vector by the summing matrix $\Svet$.

More precisely, denoting $\widehat{a}_1$ and $\widehat{\bvet}$ the base forecasts of, respectively, the total (top level) and the $n_b$ bottom time series for a fixed forecast horizon, for a given $(n_b \times 1)$ vector $\pvet$ of combination weights, i.e. $0 < p_i < 1$, $\displaystyle\sum_{i=1}^{n_b}p_i = 1$, \cite{Hollyman2021} show that the Level-1 Conditional Coherent ($L_1CC$) bts reconciled forecasts are given by
\begin{equation}
\label{LCC1}
\widetilde{b}_i^{(1)} = \widehat{b}_i + p_i\left(\widehat{a}_1 - \displaystyle\sum_{j=1}^{n_b}\widehat{b}_j\right), \quad i=1,\ldots,n_b,
\end{equation}
and the complete reconciled forecasts vector is given by $\widetilde{\yvet}^{(1)} = \Svet \widetilde{\bvet}^{(1)}$.
Furthermore, by re-stating expression (\ref{LCC1}) as
\begin{equation}
	\label{LCC1fc}
	\widetilde{b}_i^{(1)} = (1 - p_i)\widehat{b}_i + p_i\left(\widehat{a}_1 - \displaystyle\sum_{\substack{j=1 \\ 
			j \ne i}}^{n_b}\widehat{b}_j\right), \quad i=1,\ldots,n_b,
\end{equation}
the $L_1CC$ reconciled forecast (\ref{LCC1fc}) can be seen as the forecast combination of the direct (base) forecast 
$\widehat{b}_i$ and of its indirect counterpart $\left(\widehat{a}_1 - \displaystyle\sum_{\substack{j=1 \\ 
		j \ne i}}^{n_b}\widehat{b}_j\right)$, which is coherent with the accounting constraint linking the total and the bottom time series, with weights given by $(1 - p_i)$ and $p_i$, respectively.
	
Expression (\ref{LCC1}) may be extended to any level $l$, $1 \le l \le L$, as 
	\begin{equation}
		\label{btildelHolly}
		\widetilde{\bvet}^{(l)} = \widehat{\bvet} + \Pvet_l\left(\widehat{\avet}_l - \Cvet_l\widehat{\bvet}\right) , \quad l =1, \ldots, L ,
	\end{equation}
where $\Pvet_l$ is the $(n_b \times n_l)$ matrix containing the combination weights of each forecasts in the $n_l$ elementary hierarchies linking each of the $n_l$ level-$l$ series to their `afferent' bts. More precisely,
\begin{equation}
	\label{Plmat}
\Pvet_l =  \begin{bmatrix}
	\pvet_{l,1}                 & \Zerovet  & \cdots & \Zerovet \\
	\Zerovet & \pvet_{l,2}                 & \cdots & \Zerovet \\
	\vdots                  & \vdots                   & \ddots & \vdots \\
	\Zerovet &  \Zerovet & \cdots & \pvet_{l,n_l}
\end{bmatrix} , \quad l=1, \ldots, L,
\end{equation}
\noindent where each $\pvet_{l,j}$, $l=1,\ldots,L$, $j=1,\ldots,n_l$,  is a vector of weights between zero and uno, which sum up to one, with a number of element equal to the bts contributing to the upper level's counterpart series.
For $l=1$ $\Pvet_l$ reduces to vector $\pvet$, and it is easy to check that $\Cvet_l\Pvet_l = \Ivet_{n_l}$, $l=1,\ldots,L$.

The algebra behind these results found by \cite{Hollyman2021} is shown in detail in appendix 1.
	In addition, \cite{Hollyman2021} show that, if the base forecasts are unbiased, the reconciled forecasts this way are unbiased as well.
	
	It should be noted that expression (\ref{LCC1}) entails an important difference as compared to the classical top-down procedure. In the latter case, independently of the top-down disaggregation variant adopted (average historical proportions, proportions of the historical averages, forecast proportions, see \citealp{Athanasopoulos2009}), if the base forecast to be disaggregated $\widehat{a}_1$  is positive - which is rather sensible in many practical situations involving a hierarchical/grouped data organization -, all the reconciled forecasts will be non-negative as well. On the other hand, expression (\ref{LCC1}) is not guaranteed to produce non-negative $\widetilde{b}_i^{(1)}$:
	if the quantity (discrepancy) $\left(\widehat{a}_1 - \displaystyle\sum_{j=1}^{n_b}\widehat{b}_j\right) < 0$, it may happen that the correction to $\widehat{b}_i$, equal to a share $0 < p_i <1$ of the discrepancy, be larger in absolute value than the base forecast to be adjusted, thus producing a negative reconciled forecast. In particular, when the discrepancy is negative and $\widehat{b}_i \le 0$, it is always $\widetilde{b}_i^{(1)} < 0$.
Therefore we argue that - when intrinsically non-negative variables are to be forecast - interpreting $L_1CC$ as an unbiased top-down reconciliation procedure leaves room for some doubts.
In the next two sections we provide a different, meaningful interpretation of expression (\ref{LCC1}), by finding the general solution to the problem of deriving reconciled forecasts according to a level conditional coherent procedure, with either exogenous or endogenous constraints, which contains the forecast reconciliation procedure by \cite{Hollyman2021} as a particular case.



\section{$LCC$ forecast reconciliation with exogenous constraints}
\label{sec:LCCexo}

Given a generic level $l$, reconciled forecasts coherent with the base forecasts of that level can be obtained by solving the following linearly constrained quadratic minimization problem:
\begin{equation}
	\label{minLCC}
\widetilde{\bvet}^{(l)}_{\text{L$_l$CC}} = \argmin_{\bvet}
\left(\bvet - \widehat{\bvet}\right)'\Wvet_b^{-1}\left(\bvet - \widehat{\bvet}\right) \quad \text{ s.t. } \Cvet_l\bvet = \widehat{\avet}_l, \quad
l=1, \ldots, L ,
\end{equation}

\noindent where $\Wvet_b$ is a $(n_b \times n_b)$ p.d. matrix, and vector $\widehat{\avet}_l$ defines the exogenous constraint to be fulfilled. The solution is given by (appendix 2):
\begin{equation}
	\label{btildel}
	\widetilde{\bvet}^{(l)}_{\text{L$_l$CC}} = \widehat{\bvet} + \Wvet_b\Cvet_l'
	\left(\Cvet_l\Wvet_b\Cvet_l'\right)^{-1}\left(\widehat{\avet}_l -\Cvet_l\widehat{\bvet}\right), \quad l=1,\ldots,L,
\end{equation}
that is
$\widetilde{\bvet}^{(l)}_{\text{L$_l$CC}} = \Lvet_l\widehat{\avet}_l + \left(\Cvet_l\Wvet_b\Cvet_l'\right)^{-1}\widehat{\bvet}$, with
$\Lvet_l = \Wvet_b\Cvet_l'
\left(\Cvet_l\Wvet_b\Cvet_l'\right)^{-1}$.

It is worth noting that formula (\ref{btildel}) is a well known expression in the field of the least squares adjustment of noisy data while fulfilling an aggregation - either cross-sectional or temporal - constraint (\citealp{Stone1942}, \citealp{Denton1971}, \citealp{ChowLin1971}, \citealp{Byron1978}). Put simply,
the reconciliation formula (\ref{btildel}) `adjusts' the bts base forecasts with a linear combination - according to the smoothing matrix $\Lvet_l$ - of the level-$l$ coherency errors $\left(\widehat{\avet}_l -\Cvet_l\widehat{\bvet}\right)$.

Matrix $\Wvet_b$ plays a crucial role in the $LCC$ reconciliation formula. From a mathematical point of view, it is the associated matrix in the quadratic objective function of problem (\ref{minLCC}). From a statistical point of view, $\Wvet_b$ is usually seen as the forecast error covariance matrix, whose evaluation/approximation is based either on simple assumptions of independence between different forecasts, which results in a diagonal $\Wvet_b$ matrix, or on using the in-sample errors of the models used to compute the base forecasts (see \citealp{Wickramasuriya2019}, \citealp{Nystrup2020}, Di Fonzo and Girolimetto, 2020, for alternative choices, respectively, in the cross-sectional, temporal, and cross-temporal forecast reconciliation frameworks).

It clearly appears the equivalence between expression (\ref{btildel}) and formula (\ref{btildelHolly}), obtained following the approach of \cite{Hollyman2021}: for $\Lvet_l = \Pvet_l$, the two expressions are equal.
This result holds when

\begin{equation}
	\label{Wb}
\Wvet_b = \begin{bmatrix}
\sigma^2_1 & \cdots & 0 & \cdots & 0 \\
	\vdots & \ddots & \vdots & \ddots & \vdots \\
	0 & \cdots & \sigma^2_i & \cdots & 0 \\
	\vdots & \ddots & \vdots & \ddots & \vdots \\
	0 & \cdots & 0 & \cdots & \sigma^2_{n_b}
\end{bmatrix} = 
 \begin{bmatrix}
	\displaystyle\frac{1}{p_1} & \cdots & 0 & \cdots & 0 \\
	\vdots & \ddots & \vdots & \ddots & \vdots \\
	0 & \cdots & \displaystyle\frac{1}{p_i} & \cdots & 0 \\
	\vdots & \ddots & \vdots & \ddots & \vdots \\
	0 & \cdots & 0 & \cdots & \displaystyle\frac{1}{p_{n_b}}
\end{bmatrix} = \left[\text{diag}\left(\pvet\right)\right]^{-1} ,
\end{equation}
that is if we assume a diagonal bts forecast error covariance matrix, where the variances on the diagonal, $\sigma^2_i$,  are equal to the the reciprocal of the relevant combination weights: $\sigma^2_i = \displaystyle\frac{1}{p_i}$, $i=1,\ldots,n_b$.

In other terms, the $L_lCC$ reconciliation formula (\ref{btildelHolly}) by \cite{Hollyman2021} may be interpreted as the solution to a linearly constrained minimization of a quadratic form with the associated diagonal matrix (\ref{Wb}). \cite{Hollyman2021} show that the diagonal pattern of matrix $\Wvet_b$ is consistent with the usual practice in forecast combination of discarding possible covariances between the forecasts to be combined (\citealp{Bates1969}). 
It should be noted that nothing prevents us to consider a full, instead of diagonal,  $\Wvet_b$ matrix. Obviously, this would pose non trivial estimation issues, which may benefit of possibly available forecast error estimates, either from in-sample residuals or out-of-sample forecast errors in validation sets. In this paper we adopt the choice by \cite{Hollyman2021} of using a simple formulation of $\Wvet_b$, consisting in the diagonal matrix of the observed variability of the bottom time series in the training set used to estimate the base forecasts, but we think that this issue is worth considering in future research on this topic.

\subsection{Combined Conditional Coherent ($CCC$) forecast reconciliation}
For any fixed level $l$, the complete vector of $L_lCC$ reconciled forecasts may be computed as 
\begin{equation}
	\label{LCCl}
\widetilde{\yvet}^{(l)} = \Svet\widetilde{\bvet}^{(l)}, \quad l=1, \ldots, L .
\end{equation}

\begin{figure}[H]
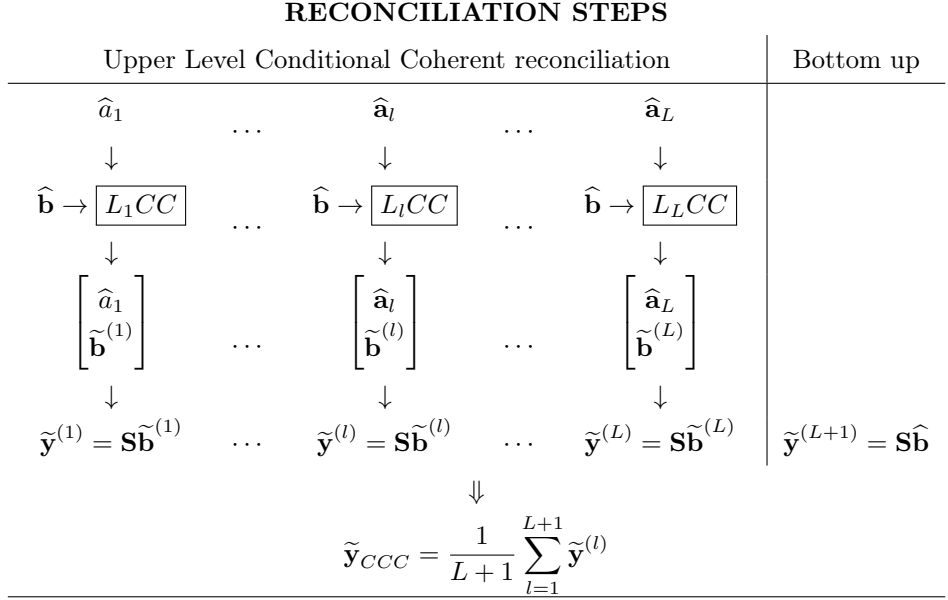

	\caption{Combined Conditional Coherent forecast reconciliation procedure with the same bts base forecasts $\widehat{\bvet}$}
	\centering
	\begin{tabular}[t]{ccccc|c}
		\multicolumn{6}{c}{\textbf{RECONCILIATION STEPS}}\\
		\multicolumn{5}{c|}{Upper Level Conditional Coherent reconciliation} & Bottom up\\
		\hline
		$\begin{array}{c} \widehat{a}_1 \\ \downarrow \end{array}$ & $\cdots$ & $\begin{array}{c} \widehat{\avet}_l \\ \downarrow \end{array}$ & $\cdots$ & $\begin{array}{c} \widehat{\avet}_L \\ \downarrow \end{array}$ \\
		$\begin{array}{c} \widehat{\bvet} \rightarrow \boxed{L_1CC} \\ \downarrow \end{array}$ & $\cdots$ & $\begin{array}{c} \widehat{\bvet} \rightarrow \boxed{L_lCC} \\ \downarrow \end{array}$ & $\cdots$ & $\begin{array}{c} \widehat{\bvet} \rightarrow \boxed{L_LCC} \\ \downarrow \end{array}$ \\
		$\begin{array}{c} \begin{bmatrix} \widehat{a}_1 \\ \widetilde{\bvet}^{(1)} \end{bmatrix} \\ \downarrow \end{array}$ & $\cdots$ &
		$\begin{array}{c} \begin{bmatrix} \widehat{\avet}_l \\ \widetilde{\bvet}^{(l)} \end{bmatrix} \\ \downarrow \end{array}$ & $\cdots$ &
		$\begin{array}{c} \begin{bmatrix} \widehat{\avet}_L \\ \widetilde{\bvet}^{(L)} \end{bmatrix} \\ \downarrow \end{array}$ \\
		$\widetilde{\yvet}^{(1)} = \Svet\widetilde{\bvet}^{(1)}$ & $\cdots$ &
		$\widetilde{\yvet}^{(l)} = \Svet\widetilde{\bvet}^{(l)}$ & $\cdots$ &
		$\widetilde{\yvet}^{(L)} = \Svet\widetilde{\bvet}^{(L)}$ & 
		$\widetilde{\yvet}^{(L+1)} = \Svet\widehat{\bvet}$\\
		\multicolumn{6}{c}{$\Downarrow$}\\
		\multicolumn{6}{c}{$\widetilde{\yvet}_{CCC}=\displaystyle\frac{1}{L+1}\displaystyle\sum_{l=1}^{L+1}\widetilde{\yvet}^{(l)}$}\\
		\hline
	\end{tabular}
	\label{fig:CCCscheme}
\end{figure}

\noindent For $l=1$, expression (\ref{LCCl}) returns reconciled forecasts obtained by summing-up the $L_1CC$ bts reconciled forecasts $\widetilde{\bvet}^{(1)}$. In the remaining cases ($1 < l \le L$), each complete vector of $L_lCC$ reconciled forecasts is the result of a middle-out reconciliation procedure applied to the $L_lCC$ bts reconciled forecasts. 
Finally, taking inspiration from \cite{Hollyman2021}, the bottom-up reconciled forecasts may be considered as the reconciled forecasts coherent with the bottom level of the hierarchy ($l=L+1$)\footnote{As we will discuss in section \ref{sec:VN525}, the empirical application of \cite{Hollyman2021} makes use of two different base forecasts when performing the $LCC$ and the bottom-up reconciliation steps, respectively.}:
$\widetilde{\yvet}^{(L+1)} = \Svet\widehat{\bvet}$.
These $L+1$ vectors contain different and coherent forecasts, so that any convex linear combination of the form
\[
\widetilde{\yvet}_{\omega} = \displaystyle\sum_{l=1}^{L+1} \omega_l \widetilde{\yvet}^{(l)} =
\Svet \displaystyle\sum_{l=1}^{L+1} \omega_l \widetilde{\bvet}^{(l)} = \Svet \widetilde{\bvet}_{\omega} ,
\]
where $\widetilde{\bvet}_{\omega} = \displaystyle\sum_{l=1}^{L+1} \omega_l \widetilde{\bvet}^{(l)}$, with $0 \le \omega_l \le 1$, $\displaystyle\sum_{l=1}^{L+1}\omega_l = 1$, is coherent as well.
Combined Conditional Coherent ($CCC$) forecasts are calculated by using equal weights, which means to combine the $L+1$ reconciled forecasts through the simple average (see Figure \ref{fig:CCCscheme})

\begin{equation}
	\label{HollyCCC}
	\widetilde{\yvet}_{CCC} = \displaystyle\frac{1}{L+1}\sum_{l=1}^{L+1} \widetilde{\yvet}^{(l)} .
\end{equation}

\noindent From this scheme it appears that the $L_lCC$ steps are logically different from the bottom-up one: in the former case, base forecasts of both upper series (part of them at each step) and bottom series, are combined through the optimization mechanism described above, while in the latter no upper time series base forecast is used. Thus, it seems rather sensible considering a variant of the $CCC$ procedure, consisting in the simple average of the `true' $L_lCC$ reconciled forecasts at the various levels of the hierarchy, that is:
\begin{equation}
	\label{CCCred}
	\widetilde{\yvet}_{LCC} = \displaystyle\frac{1}{L}\displaystyle\sum_{l=1}^{L}\widetilde{\yvet}^{(l)} .
\end{equation}

\subsection{Some examples}
\label{sec:examples}
\subsection*{Level 1 Conditional Coherent ($L_1CC$) reconciliation}
Consider $l=1$, and let $\Wvet_b$ be defined as in (\ref{Wb}). 
Since $\Cvet_1 = {\bf 1}_{n_b}'$, 
it is immediately recognized that
$
\Wvet_b\Cvet_1' = \begin{bmatrix}
	\sigma^{2}_{1} \; \ldots \;
	\sigma^{2}_{i} \; \ldots \;
	\sigma^{2}_{n_b}
\end{bmatrix}'
$, and
$\Cvet_1\Wvet_b\Cvet_1' = \displaystyle\sum_{j=1}^{n_b}\sigma^{2}_j$.
The generic item of the vector obtained through the reconciliation formula (\ref{btildel}) is thus given by:
\[
\widetilde{b}^{(1)}_i = \widehat{b}_i + \displaystyle\frac{\sigma^{2}_i}{\displaystyle\sum_{j=1}^{n_b}\sigma^{2}_j}
\left(\widehat{a}_1 - \displaystyle\sum_{j=1}^{n_b}\widehat{b_j}\right) , \quad i=1,\ldots, n_b ,
\]
that is
\[
\widetilde{b}^{(1)}_i = \left(\displaystyle\frac{\displaystyle\sum_{\substack{j=1 \\ j \ne i}}^{n_b}\sigma^{2}_j}{\displaystyle\sum_{j=1}^{n_b}\sigma^{2}_j}\right) \widehat{b}_i +
\displaystyle\frac{\sigma^{2}_i}{\displaystyle\sum_{j=1}^{n_b}\sigma^{2}_j}
\left(\widehat{a}_1 - 
\sum_{\substack{j=1 \\ j \ne i}}^{n_b} \widehat{b_j}\right) , \quad i=1,\ldots, n_b ,
\]
which corresponds to equation (\ref{LCC1fc}) in \cite{Hollyman2021}.

\subsection*{CCC reconciliation in a three-level hierarchy}
When a level $l>1$ is considered, again with a diagonal $\Wvet_b$, the scalar expressions of the $L_lCC$ reconciled forecasts are more complex. Nevertheless, starting from the simple hierarchy in figure \ref{Fig:hts1} (which corresponds to the example 2 in \citealp{Hollyman2021}), we can easily interpret the final result. In this case it is easy to check that
\[
\Wvet_b\Cvet_2' = \begin{bmatrix}
	\sigma^{2}_A & 0 \\
	\sigma^{2}_B & 0 \\
	0 & \sigma^{2}_C \\
	0 & \sigma^{2}_D \\
	0 & \sigma^{2}_E	
\end{bmatrix}
\quad \text{ and } \quad
\Cvet_2 \Wvet_b \Cvet_2' =
\begin{bmatrix}
	\sigma^{2}_A + \sigma^{2}_B & 0 \\
	0 & \sigma^{2}_C + \sigma^{2}_D + \sigma^{2}_E
\end{bmatrix} ,
\]
from which we obtain
\[
\widetilde{b}^{(2)}_i = \left\{ \begin{array}{ll}
	\widehat{b}_i + \displaystyle\frac{\sigma^{2}_i}{\sigma^{2}_A + \sigma^{2}_B}
	\left(\widehat{a}_X - \widehat{b}_A - \widehat{b}_B\right) & i=A,B \\[.5cm]
	\widehat{b}_i + \displaystyle\frac{\sigma^{2}_i}{\sigma^{2}_C + \sigma^{2}_D + \sigma^{2}_E}
	\left(\widehat{a}_Y - \widehat{b}_C - \widehat{b}_D - \widehat{b}_E\right) & i=C,D,E
\end{array}
\right. ,
\]
where $\widehat{a}_X$ and $\widehat{a}_Y$
denote the base forecasts of $X$ and $Y$, respectively.

Put in other words, level 2, consisting of series $X$ and $Y$, is `decomposed' in 2 elementary hierarchies, each consisting in a single aggregated series belonging to that level, and in the corresponding bts. Then, a $L_1CC$ forecast reconciliation procedure is applied to each elementary hierarchy.

The whole reconciliation is obtained by summing-up the reconciled forecasts of the bts (and thus the final result should be viewed as a middle-out forecast reconciliation):
\[
\widetilde{a}_T^{(2)} = \widehat{a}_X + \widehat{a}_Y =
\widetilde{b}^{(2)}_A + \widetilde{b}^{(2)}_B + \widetilde{b}^{(2)}_C +
\widetilde{b}^{(2)}_D + \widetilde{b}^{(2)}_E .
\]
Notice that in general $\widetilde{a}^{(2)}_T \ne \widehat{a}_T$, and likewise all the reconciled forecasts this way are different from those obtained if we consider the reconciliation conditional to level 1 (that is, to the base forecast of the top-level series, $\widehat{a}_T$).
As for the $CCC$ reconciled forecasts, by using expression (\ref{HollyCCC}) we get:
\begingroup\footnotesize
\[
\begin{array}{rcl}
	\widetilde{a}_{T,CCC} & = & \displaystyle\frac{1}{3}\left(\widetilde{a}_T^{(1)} + \widetilde{a}_T^{(2)} + \widetilde{a}_T^{(3)}\right) = \displaystyle\frac{1}{3}\left(\widehat{a}_T +
	\underbrace{\widehat{a}_X + \widehat{a}_Y}_{\widetilde{a}_T^{(2)}} +
	\underbrace{\widehat{b}_A + \widehat{b}_B + \widehat{b}_C +\widehat{b}_D + \widehat{b}_E}_{\widetilde{a}_T^{(3)}}	
	\right) \\[.5cm]
	\widetilde{a}_{X,CCC} & = & \displaystyle\frac{1}{3}\left(\widetilde{a}_X^{(1)} + \widetilde{a}_X^{(2)} + \widetilde{a}_X^{(3)}\right) = \displaystyle\frac{1}{3}\left(
	\underbrace{\widetilde{b}_A^{(1)} + \widetilde{b}_B^{(1)}}_{\widetilde{a}_X^{(1)}} +
	\widehat{a}_X +
	\underbrace{\widehat{b}_A + \widehat{b}_B}_{\widetilde{a}_X^{(3)}}
	\right) \\[.5cm]
	\widetilde{a}_{Y,CCC} & = & \displaystyle\frac{1}{3}\left(\widetilde{a}_Y^{(1)} + \widetilde{a}_Y^{(2)} + \widetilde{a}_Y^{(3)}\right) = \displaystyle\frac{1}{3}\left(
	\underbrace{\widetilde{b}_C^{(1)} + \widetilde{b}_D^{(1)} + \widetilde{b}_E^{(1)}}_{\widetilde{a}_Y^{(1)}}
	+ \widehat{a}_Y + 
	\underbrace{\widehat{b}_C + \widehat{b}_D + \widehat{b}_E}_{\widetilde{a}_Y^{(3)}}
	\right)\\[.5cm]
	\widetilde{b}_{A,CCC} & = & \displaystyle\frac{1}{3}\left(\widetilde{b}_A^{(1)} + \widetilde{b}_A^{(2)} + \widehat{b}_A\right) \\[.5cm]
	\widetilde{b}_{B,CCC} & = & \displaystyle\frac{1}{3}\left(\widetilde{b}_B^{(1)} + \widetilde{b}_B^{(2)} + \widehat{b}_B\right) \\[.5cm]
	\widetilde{b}_{C,CCC} & = & \displaystyle\frac{1}{3}\left(\widetilde{b}_C^{(1)} + \widetilde{b}_C^{(2)} + \widehat{b}_C\right) \\[.5cm]
	\widetilde{b}_{D,CCC} & = & \displaystyle\frac{1}{3}\left(\widetilde{b}_D^{(1)} + \widetilde{b}_D^{(2)} + \widehat{b}_D\right) \\[.5cm]
	\widetilde{b}_{E,CCC} & = & \displaystyle\frac{1}{3}\left(\widetilde{b}_E^{(1)} + \widetilde{b}_E^{(2)} + \widehat{b}_E\right) \\[.5cm]
\end{array}
\]\endgroup
It is worth noting that the top-level $CCC$ reconciled forecast, $\widetilde{a}_{T,CCC}$, is given by the simple average of the direct base forecast $\widehat{a}_T$, and of the two indirect forecasts obtained by summing the base forecasts at the intermediate ($\widetilde{a}_T^{(2)} = \widehat{a}_X + \widehat{a}_Y$), and at the bottom
($\widetilde{a}_T^{(3)} = \widehat{b}_A + \widehat{b}_B + \widehat{b}_C + \widehat{b}_D + \widehat{b}_E$) levels, respectively, without any use of the uncertainty associated to the bts base forecasts. This information is instead taken into account when computing the $CCC$ reconciled forecasts of variables $X$ and $Y$, through the $L_1CC$ reconciled forecasts
$\widetilde{\bvet}^{(1)} = \left[\widetilde{b}_A^{(1)} \; \; \widetilde{b}_B^{(1)} \; \; \widetilde{b}_C^{(1)} \; \; \widetilde{b}_D^{(1)} \; \; \widetilde{b}_E^{(1)} \right]'$.

\section{$LCC$ forecast reconciliation with endogenous constraints}
\label{sec:LCCendo}
In summary, the approach by \cite{Hollyman2021} consists in (i) decomposing the hierarchical/grouped structure in a sequence of elementary hierarchies, and (ii) for each elementary hierarchy, the bts base forecasts are reconciled according to a \emph{level conditional coherent} procedure with exogenous constraints. Furthermore, in the approach suggested by \cite{Hollyman2021}, only weights given by the variances of the base forecasts are considered, discarding - as it is often found in the forecast combination literature (\citealp{Bates1969}) - the covariances between couples of forecasts.

In the following we broaden the perspective, by relaxing the assumption of exogenous constraints in the optimization program (\ref{minLCC}). We consider a level conditional reconciliation procedure with endogenous constraints, which means that level-$l$ reconciled forecasts are computed by transforming the base forecasts looking for internal coherency of the targets, without imposing the external constraint of the uts base forecasts, as in the approach described in section \ref{sec:LCCexo}.

Denote with $\yvet_l = \begin{bmatrix}
	\avet_l \\ \bvet
\end{bmatrix}$ the $\left[(n_b+n_l) \times 1\right]$ vector of the target forecasts of the level-$l$ series, $\avet_l$, and of the bts, $\bvet$. The corresponding base forecasts form the vector $\widehat{\yvet}_l = \begin{bmatrix}
	\widehat{\avet}_l \\ \widehat{\bvet}
\end{bmatrix}$.
Let $\Uvet_l = \left[\Ivet_{n_l} \; \; -\Cvet_l\right]$ be the $\left[n_l \times (n_b + n_l)\right]$ matrix of the homogeneous constraints valid for the level-$l$ series:
\[
\Uvet_l' \yvet_l = \Zerovet_{n_l}, \quad l=1,\ldots, L .
\]
Notice that in this case the cross-sectional aggregation constraint is endogenous, and is valid for the level-$l$ aggregated series as well, whose base forecasts $\widehat{\avet}_l$, unlike the procedure by \cite{Hollyman2021}, are not considered as exogenous constraints for the bts reconciled forecasts, but are themselves object of the reconciliation process. In order this may happen, we need to have weights for the base forecasts of the aggregated series. Thus, let $\Wvet_l$ be the $\left[(n_l+n_b) \times (n_l+n_b))\right]$ matrix (assumed diagonal) containing the variances of the $n_l + n_b$ base forecasts to be reconciled in coherence with the linear aggregation relationships between the bts and the level-$l$ aggregated series. In this case the $n_l + n_b$ level-$l$ reconciled forecasts are given by (\citealp{DiFonzoGiro2020}):
\begin{equation}
	\label{ytildek}
	\widetilde{\yvet}_l = \Mvet_l\widehat{\yvet}, \quad
	\text{ with } \quad \Mvet_l = \Ivet_{n_l+n_b} - \Wvet_l\Uvet_l\left(\Uvet_l'\Wvet_l\Uvet_l\right)^{-1}\Uvet_l' , \quad l=1,\ldots,L.
\end{equation}
In order to express the complete $(n \times 1)$ vector of reconciled forecasts $\widetilde{\yvet}^{(l)}_{\text{en}}$, where the superscript $^{(l)}$ stands for the level at which the reconciliation is performed, and the subscript `$_{\text{en}}$' stands for `endogenously constrained', it is sufficient to apply the structural sum matrix $\Svet$ to the vector formed by the bottom $n_b$ items of vector $\widetilde{\yvet}_l$, that is the vector $\widetilde{\bvet}^{(l)}_{\text{en}}$ of the bts reconciled forecasts\footnote{Formally, let  $\Jvet_l' = \begin{bmatrix}
		\Zerovet_{n_b \times n_l} & \Ivet_{n_b}
	\end{bmatrix}$ be the $\left[n_b \times \left(n_l+n_b\right)\right]$ matrix which `extracts' the last $n_b$ rows from a $\left[\left(n_l+n_b\right) \times 1 \right]$ vector, it is $\widetilde{\bvet}^{(l)}_{\text{en}} = \Jvet_l'\widetilde{\yvet}_l$, $l=1,\ldots,L$.}:
$\widetilde{\yvet}^{(l)}_{\text{en}} = \Svet\widetilde{\bvet}^{(l)}_{\text{en}}$,
$l=1,\ldots, L$.

It is instructive to consider what happens when reconciliating a very simple hierarchy of 3 series, with $T = X + Y$. In this case $l=1$, $n_l=1$, $n_b=2$ and
\[\Wvet_1 = \begin{bmatrix}
	\sigma^2_T & 0 & 0 \\
	0 & \sigma^2_X & 0 \\
	0 & 0 & \sigma^2_Y
\end{bmatrix} , \quad
\Uvet_1' = \begin{bmatrix}
	1 & -1 & -1
\end{bmatrix}, \quad \Wvet_1\Uvet_1 = \begin{bmatrix}
	\sigma^2_T \\ - \sigma^2_X \\ -\sigma^2_Y
\end{bmatrix}, \quad \Uvet_1'\Wvet_1\Uvet_1 = \sigma^2_T+\sigma^2_X+\sigma^2_Y .
\]
After a bit of algebra, it is found that the reconciled forecasts are given by:
\[
\begin{array}{rcl}
	\widetilde{T}_{\text{en}} & = & \widehat{T} - \displaystyle\frac{\sigma^2_T}{\sigma^2_T + \sigma^2_X+\sigma^2_Y}\left(\widehat{T} - \widehat{X} - \widehat{Y}\right) \\[.5cm]
	\widetilde{X}_{\text{en}} & = & \widehat{X} + \displaystyle\frac{\sigma^2_X}{\sigma^2_T+\sigma^2_X+\sigma^2_Y}\left(\widehat{T} - \widehat{X} - \widehat{Y}\right) \\[.5cm]
	\widetilde{Y}_{\text{en}} & = & \widehat{Y} + \displaystyle\frac{\sigma^2_Y}{\sigma^2_T+\sigma^2_X+\sigma^2_Y}\left(\widehat{T} - \widehat{X} - \widehat{Y}\right)
\end{array} .
\]
This result has the usual `linearly constrained least squares adjustment of noisy data' interpretation (\citealp{Stone1942}, \citealp{Byron1978}), where the reconciled forecast is given by the algebraic sum of the base forecast and of a share of the discrepancy observed in the base forecasts, where the share is proportional to the variance (uncertainty) of the forecast. Furthermore, in this case too it is possible to give the expressions above a forecast combination interpretation, as in \cite{Hollyman2021}.
For, each reconciled forecast, at both upper and bottom level, can be written as the combination of the `direct' (i.e., base) forecast, and the `indirect' (i.e., implicitly obtained using the accounting relationships) one:
\[
\begin{array}{rcl}
	\widetilde{T}_{\text{en}} & = & \left(\displaystyle\frac{\sigma^2_X+\sigma^2_Y}{\sigma^2_T+\sigma^2_X+\sigma^2_Y}\right)\widehat{T} + \left(\displaystyle\frac{\sigma^2_T}{\sigma^2_T+\sigma^2_X+\sigma^2_Y}\right)
	\left(\widehat{X} + \widehat{Y}\right) \\[.5cm]
	\widetilde{X}_{\text{en}} & = &  \left(\displaystyle\frac{\sigma^2_T+\sigma^2_Y}{\sigma^2_T+\sigma^2_X+\sigma^2_Y}\right)\widehat{X} + \left(\displaystyle\frac{\sigma^2_X}{\sigma^2_T+\sigma^2_X+\sigma^2_Y}\right)
	\left(\widehat{T} - \widehat{Y}\right) \\[.5cm]
	\widetilde{Y}_{\text{en}} & = & \left(\displaystyle\frac{\sigma^2_T+\sigma^2_X}{\sigma^2_T+\sigma^2_X+\sigma^2_Y}\right)\widehat{Y} + \left(\displaystyle\frac{\sigma^2_Y}{\sigma^2_T+\sigma^2_X+\sigma^2_Y}\right)
	\left(\widehat{T} - \widehat{X}\right)
\end{array} .
\]
However, unlike what happens in the $L_lCC$ forecast reconciliation with exogenous constraints, in this case all the base forecasts are `revised' in view of all the variances, not only the bts ones.
The development of an analogous result for the \textit{toy example} considered in Figure \ref{Fig:hts1}, can be found in appendix 3.


\section{Empirical application}
\label{sec:VN525}
In this section, we reprise and extend the forecasting experiment performed by \cite{Hollyman2021} on the Australian Tourism Demand (\textit{Visitor Nights}, VN) dataset.
 We start by (i) re-assessing the results found by \cite{Hollyman2021} using the forecast accuracy evaluation approach recommended by \cite{Davydenko2013}, and (ii) considering some non-negativity issues that emerge during both the base forecasting and the reconciliation phases of the analysis. On this latter point we note that, with the notably exceptions of \cite{Wickramasuriya2020} and \cite{Kourentzes2021}, this issue is generally overlooked in the forecast reconciliation literature, even though it has not irrilevant implications as for the interpretation of the results (e.g., a negative forecast touristic demand makes no sense). Possible, though not fully convincing, motivations for this are that, on the practical side, adopting non-negative forecast reconciliation procedures is perceived as computation burdensome, and on the theorethical side, assuring non-negativity does not preserve unbiasedness in the final non-negative reconciled forecasts (on this point, see \citealp{BenTaieb2019}, \citealp{Wickramasuriya2020}, \citealp{Wickramasuriya2021b}).

\subsection{Performance measures for multiple comparisons}

We evaluate the performance of multiple (say, $J>1$) forecast reconciliation approaches through accuracy indices
calculated on the forecast error
$$
\hat{e}_{i,h,t}^{j} = y_{i,t+h} - \hat{y}_{i,h,t}^{j}, \quad
	i=1,\ldots,525, \quad
	h=1,\ldots,12 , \quad
	j=0,\ldots,J, \quad
    t=1,\ldots, q_h , 
$$
where $y$ and $\hat{y}$ are the observed and forecast values, respectively, $i$ denotes the series ($i=1,\ldots,221$, for the uts, $i=222,\ldots,525$, for the bts), $h$ denotes the forecast horizon, $t$ is the forecast origin ($t=1$ corresponds to 2005:12, $q_1=132, \ldots, q_{12}=121$), and $j=0$ denotes the automatic ETS base forecasts.
The accuracy across multiple series and forecast horizons is evaluated following \cite{Davydenko2013} (see also \citealp{Kourentzes2020}), who recommend the use of a metric based on aggregating performance ratios across time series using the weighted geometric mean. We consider both the Average Relative Mean Absolute Error (AvgRelMAE), and
the Average Relative Mean Square Error (AvgRelMSE), obtained by transforming MAE and MSE index, respectively. As the conclusions drawn from both indices are basically the same, for space reason in the rest of the paper we comment only on the AvgRelMSE index\footnote{The interested reader may find all the tables and graphs based on the AvgRelMAE index in the on-line appendix.}.

For a fixed series $i$, forecast origin $h$, and approach $j$, the MSE index is given by the average across all $q_h$ forecast origins of the squared forecast errors:
\begin{equation}
	\label{MSEijh}
	\text{MSE}_{i,h}^{j} = \frac{1}{q_h} \displaystyle\sum_{t=1}^{q_h}\left(\hat{e}_{i,h,t}^{j}\right)^2 , \quad
		i=1,\ldots,525, \quad h=1,\ldots,12 , \quad
		j=0,\ldots,J.
\end{equation}


\noindent The AvgRelMSE of a forecasting approach for a given horizon is the geometric mean across all 525 series of the MSE ratio 
over a benchmark given by the base, incoherent ETS forecasts:
\begin{equation}
	\label{AvgRelMSEhj}
	\text{AvgRelMSE}_{h}^{j} =
	\left(\displaystyle\prod_{i=1}^{525} \text{rMSE}_{i,h}^{j}\right)^{\frac{1}{525}} , \; h=1,\ldots,12 , \; j=0,\ldots,J,
\end{equation}
where $\text{rMSE}_{i,h}^{j} = \displaystyle\frac{\text{MSE}_{i,h}^{j}}{\text{MSE}_{i,h}^{0}}$ is the relative MSE.
If a forecast outperforms the base forecasts, then the AvgRelMSE becomes smaller than one and vice-versa, and the percentage improvement in accuracy over the benchmark can be calculated as $\left(1-\text{AvgRelMSE}_{h}^{j}\right) \times 100$.
Expression (\ref{AvgRelMSEhj}), which refers to all 525 time series, can be re-stated for (i) groups of variables (e.g., bts and uts), and (ii) multiple forecast horizons (e.g., $h=$ 1:6, $h=$ 1:12).
Furthermore, we use the non-parametric Friedman and the post-hoc `Multiple Comparison with the Best' (MCB) Nemenyi tests (\citealp{Koning2005}; 
\citealp{Kourentzes2019}; \citealp{Makridakis2020}) 
to establish if the forecasting performances of the considered approaches are significantly different.



\subsection{The data and the original forecasting experiment of \cite{Hollyman2021}}

We consider 228 monthly observations (Jan 1998 - Dec 2016) of the Australiam touristic flows (\textit{Visitor Nights}) measured by the public project ``National Visitor Survey'' (\citealp{Wickramasuriya2019}).
The time series dataset consists in a grouped time series obtained by combination of a hierarchy by geographical division (destination) with a classification by Purpose of Travel (PoT). The Australian total is thus disaggregated by States (7), Zones (27), and Regions (76). Nominally, the geographic hierarchy comprises 111 destinations. However, since 6 Zones 
consist of a single Region, the non-redundant knots of the hierarchy are 105 instead of 111. Thus, we face an `unbalanced hierarchy' (see sec. \ref{sec:LCCexo}).
As PoT can assume 4 different values: holiday (Hol), visiting friends and relatives (Vis), business (Bus), and other (Oth), in the grouped series obtained by crossing geographic divisions and PoT, 24 knots (6 Zones with a single region by each PoT category) are redundant. The 304 most disaggregated variables, when combined according to the considered classifications, and discarding the duplications, produce 221 upper time series.
In summary, the 
dataset comprises 8 levels, 
with 304 bottom time series, and 221 upper time series (525 \textit{unique} time series in all)\footnote{Details can be found in the on-line appendix.}.


\cite{Hollyman2021} have performed a rolling forecast experiment with fixed length (96 months) window, producing base forecasts with forecast horizons varying from 1 to 12 months. Exponential Smoothing models (ETS, \citealp{Hyndman2008}) have been used to compute the base forecasts of the 525 series according to the automatic default of the \texttt{R} package \texttt{forecast} (\citealp{forecast2021})\footnote{We acknowledge Ross Hollyman, who kindly made us available the Python scripts used in \cite{Hollyman2021}, thus allowing us to fully reproduce their results through the \texttt{R} package \texttt{FoReco} (\citealp{FoReco2021}).}.
\cite{Hollyman2021} claim that, in agreement with \cite{Wickramasuriya2019}, ETS models produce ``substantially more accurate base forecasts than ARIMA based models in this setting, therefore presenting a more challenging environment for forecast combination techniques we consider''. We basically agree on this point, stressing however that the intrinsic non-negative nature of the variables under analysis would have been better taken into account by modeling the log-transformed data, possibly adopting the same strategy as in \cite{Wickramasuriya2020} for dealing with the null values present in the observations. Anyway, the focus of the paper being on the insights of the proposal by \cite{Hollyman2021}, and the potential of the new reconciliation approach, we decided to continue using base forecasts obtained by ETS models in the level of the variables, limiting ourselves to recognize and quantitatively assess 
the problem during the base forecasting phase of the experiment, postponing the non-negativity issues to the reconciliation phase, where possibly negative base forecasts will be reconciled through effective non-negative linearly constrained least squares procedures (\citealp{OSQP2019}, \citealp{Stellato2020}, \citealp{Wickramasuriya2020}, \citealp{Hyndman2020},  \citealp{FoReco2021}).

\begin{figure}[H]
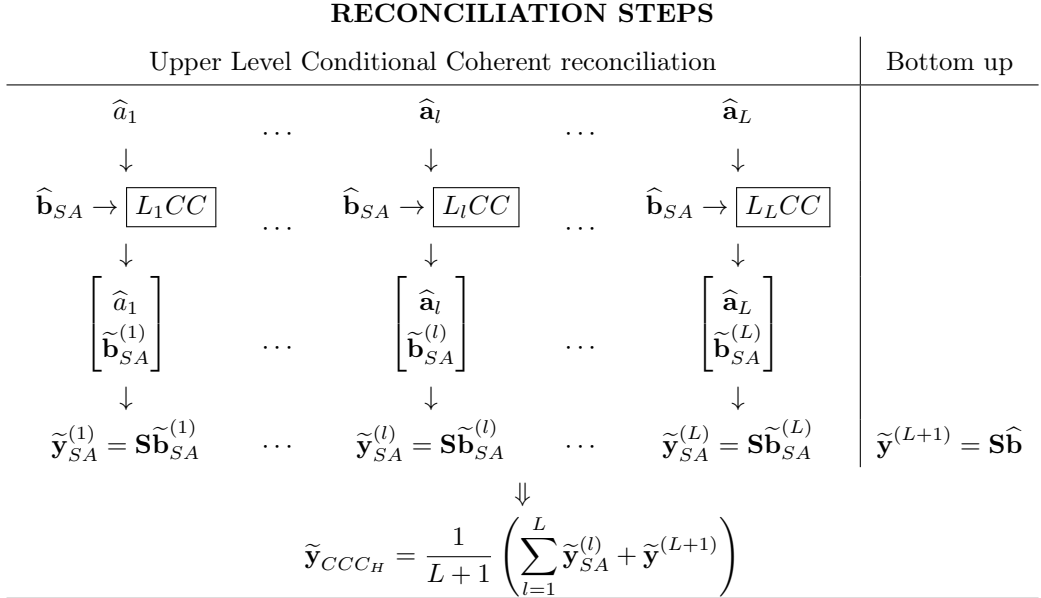

	\caption{$CCC_H$: Combined Conditional Coherent forecast reconciliation procedure according to Hollyman et al. (2021). In the Upper Level Conditional Coherent reconciliation steps the base forecasts $\widehat{\bvet}_{SA}$ are used, while in the bottom-up reconciliation $\widehat{\bvet}$ (automatic ETS) is used.}
	\centering
	\begin{tabular}[t]{ccccc|c}
		\multicolumn{6}{c}{\textbf{RECONCILIATION STEPS}}\\
		\multicolumn{5}{c|}{Upper Level Conditional Coherent reconciliation} & Bottom up\\
		\hline
		$\begin{array}{c} \widehat{a}_1 \\ \downarrow \end{array}$ & $\cdots$ & $\begin{array}{c} \widehat{\avet}_l \\ \downarrow \end{array}$ & $\cdots$ & $\begin{array}{c} \widehat{\avet}_L \\ \downarrow \end{array}$ \\
		$\begin{array}{c} \widehat{\bvet}_{SA} \rightarrow \boxed{L_1CC} \\ \downarrow \end{array}$ & $\cdots$ & $\begin{array}{c} \widehat{\bvet}_{SA} \rightarrow \boxed{L_lCC} \\ \downarrow \end{array}$ & $\cdots$ & $\begin{array}{c} \widehat{\bvet}_{SA} \rightarrow \boxed{L_LCC} \\ \downarrow \end{array}$ \\
		$\begin{array}{c} \begin{bmatrix} \widehat{a}_1 \\ \widetilde{\bvet}^{(1)}_{SA} \end{bmatrix} \\ \downarrow \end{array}$ & $\cdots$ &
		$\begin{array}{c} \begin{bmatrix} \widehat{\avet}_l \\ \widetilde{\bvet}^{(l)}_{SA} \end{bmatrix} \\ \downarrow \end{array}$ & $\cdots$ &
		$\begin{array}{c} \begin{bmatrix} \widehat{\avet}_L \\ \widetilde{\bvet}^{(L)}_{SA} \end{bmatrix} \\ \downarrow \end{array}$ \\
		$\widetilde{\yvet}^{(1)}_{SA} = \Svet\widetilde{\bvet}^{(1)}_{SA}$ & $\cdots$ &
		$\widetilde{\yvet}^{(l)}_{SA} = \Svet\widetilde{\bvet}^{(l)}_{SA}$ & $\cdots$ &
		$\widetilde{\yvet}^{(L)}_{SA} = \Svet\widetilde{\bvet}^{(L)}_{SA}$ & 
		$\widetilde{\yvet}^{(L+1)} = \Svet\widehat{\bvet}$\\
		\multicolumn{6}{c}{$\Downarrow$}\\
		\multicolumn{6}{c}{$\widetilde{\yvet}_{CCC_H}=\displaystyle\frac{1}{L+1}\left(\displaystyle\sum_{l=1}^{L}\widetilde{\yvet}^{(l)}_{SA} + \widetilde{\yvet}^{(L+1)}\right)$}\\
		\hline
	\end{tabular}
	\label{fig:CCCHscheme}
\end{figure}

Besides the (incoherent) base forecasts, \cite{Hollyman2021} have considered the reconciled forecasts produced by the following approaches\footnote{For homogeneity, we use the labels of the reconciliation approaches adopted in the rest of the paper. The correspondence with those originally used by \cite{Hollyman2021} is as follows: $L_1CC \equiv$ TD, $wls \equiv$ WLSv, $shr \equiv$ MinTShrink, and $CCC_H \equiv$ CCC.}:
\begin{itemize}
\item Bottom Up (BU): forecasts obtained by simple summation of the automatic ETS base forecasts for the 304 most disaggregate series. 
\item Level 1 Coherent Combination ($L_1CC$): coherent forecasts with the base forecast of the top-level of the hierarchy (Total Australia), computed using expression (\ref{LCC1}).
\item Top Down Historical Proportions (TDHP): forecasts obtained through a top-down reconciliation procedure using the ‘HP2’ approach of \cite{Athanasopoulos2009}, where the historical proportions disaggregation coefficients are computed on monthly basis. 
\item OLS: the original forecast reconciliation model of \cite{Hyndman2011}, which assumes that the base forecasts are uncorrelated and identically distributed. 
It can be seen as a particular case of the Minimum Trace reconciliation approach of \cite{Wickramasuriya2019}.
\item $wls$: like OLS, with the error variances of each series taken into account; the forecast errors are assumed to be uncorrelated but heteroskedastic (differing variances).
\item $shr$: the Minimum Trace optimal approach of \cite{Wickramasuriya2019} based on a shrinkage estimator of the covariance matrix 
of forecast errors.
\item $CCC_H$: an equally weighted average of 8 forecasts for each series derived from the 8 levels of the VN525 hierarchy (see Figure \ref{fig:CCCHscheme}). For the first 7 levels, the ETS base forecasts of the upper time series are used in the $L_lCC$, $l=1,\ldots, 7$, reconciliation formula (\ref{btildel}), with the base forecasts given by the seasonal averages of the observations in the training set used to estimate the models (i.e., the preceding 96 months), which we denote by the $(304 \times 1)$ vector $\widehat{\bvet}_{SA}$.
Matrix $\Wvet_b$ in (\ref{btildel}) is diagonal, with non-zero entries equal to the variance of each bottom time series measured over the training set on a seasonal basis (details in \citealp{Hollyman2021}). 
The last term in the equally weighted average is given by the BU reconciled forecasts considered so far.
\end{itemize}

In most replications of both base forecasting and reconciliation phases of the experiment, negative forecasts have been produced. In 114 out of 132 replications of the experiment, and for a maximum of 8 series in the same replication, at least one automatic ETS base forecast was negative. The reconciliation phase seems to somehow worsen this issue: the OLS approach always produces a few negative reconciled forecasts (ranging from 15 to 119 series at each replication), while this phenomenon, though still not negligible, is less present in the remaining cases. It is worth noting that the $L_1CC$ approach gives negative forecasts in more than 63\% of the replications (84 out of 132), up to a maximum of about 9\% of series (47 out of 525).
As expected, negative forecasts are mostly present at the most disaggregate level (L8: Regions by PoT), with a less pronounced intensity of the phenomenon for the upper levels of the hierarchy (details can be found in the on-line appendix).
In order to guarantee comparability with the results of \cite{Hollyman2021}, Table \ref{tab:HollyTableMSE} 
shows the AvgRelMSE's for the approaches considered in their paper, without any treatment of the negative values produced by the forecasting experiment.

\begin{table}[H]
\caption{\label{tab:HollyTableMSE}Monthly forecasts reconciliation
	in the forecasting experiment on the Australian tourism dataset: \textbf{AvgRelMSE} of the approaches considered by Hollyman et al. (2021). Approach TDHP apart, some reconciled forecasts are negative. 
	Bold entries identify the best performing approaches. Red entries identify the approaches worsening the automatic ETS base forecasts' accuracy.}
\centering
\begingroup

\footnotesize
\setlength\tabcolsep{4pt}
\begin{tabular}[t]{lccccccc}
\hline
 & \multicolumn{7}{c}{Forecast horizon}\\
Approach & 1 & 2 & 3 & 6 & 12 & 1:6 & 1:12\\
\hline
\multicolumn{8}{c}{\textit{all (525 series)}} \\
BU & 0.9974 & 0.99342 & 0.9924 & 0.9976 & {\color{red}1.0010} & 0.9956 & 0.9985\\
$L_1CC$ & {\color{red}1.0032} & {\color{red}1.0041} & {\color{red}1.0031} & 0.9966 & 0.9797 & {\color{red}1.0004} & 0.9935\\
TDHP & {\color{red}1.0055} & {\color{red}1.0070} & {\color{red}1.0059} & 0.9980 & 0.9785 & {\color{red}1.0027} & 0.9944\\
OLS & {\color{red}1.0740} & {\color{red}1.0748} & {\color{red}1.0781} & {\color{red}1.0730} & {\color{red}1.0991} & {\color{red}1.0757} & {\color{red}1.0790}\\
$wls$ & 0.9806 & 0.9805 & 0.9809 & 0.9816 & 0.9837 & 0.9809 & 0.9818\\
$shr$ & \textbf{0.9745} & \textbf{0.9761} & 0.9760 & 0.9758 & 0.9783 & 0.9757 & 0.9763\\
$CCC_H$ & 0.9764 & 0.9765 & \textbf{0.9759} & \textbf{0.9726} & \textbf{0.9664} & \textbf{0.9743} & \textbf{0.9713}\\
\multicolumn{8}{c}{\textit{upper time series (221 series)}} \\
BU & 0.9939 & 0.9863 & 0.9821 & 0.9943 & {\color{red}1.0024} & 0.9897 & 0.9965\\
$L_1CC$ & 0.9769 & 0.9790 & 0.9782 & 0.9753 & 0.9548 & 0.9765 & 0.9705\\
TDHP & 0.9801 & 0.9826 & 0.9815 & 0.9768 & 0.9529 & 0.9792 & 0.9714\\
OLS & 0.9997 & 0.9984 & {\color{red}1.0025} & 0.9986 & {\color{red}1.0141} & 1.0000 & {\color{red}1.0015}\\
$wls$ & 0.9614 & 0.9595 & 0.9595 & 0.9626 & 0.9676 & 0.9609 & 0.9633\\
$shr$ & 0.9537 & 0.9542 & 0.9534 & 0.9554 & 0.9609 & 0.9545 & 0.9563\\
$CCC_H$ & \textbf{0.9476} & \textbf{0.9476} & \textbf{0.9470} & \textbf{0.9486} & \textbf{0.9451} & \textbf{0.9471} & \textbf{0.9468}\\
\multicolumn{8}{c}{\textit{bottom time series (304 series)}} \\
BU & 1.0000 & 1.0000 & 1.0000 & 1.0000 & 1.0000 & 1.0000 & 1.0000\\
$L_1CC$ & {\color{red}1.0227} & {\color{red}1.0227} & {\color{red}1.0217} & {\color{red}1.0124} & 0.9982 & {\color{red}1.0181} & {\color{red}1.0105}\\
TDHP & {\color{red}1.0244} & {\color{red}1.0251} & {\color{red}1.0240} & {\color{red}1.0138} & 0.9976 & {\color{red}1.0201} & {\color{red}1.0115}\\
OLS & {\color{red}1.1315} & {\color{red}1.1339} & {\color{red}1.1366} & {\color{red}1.1306} & {\color{red}1.1653} & {\color{red}1.1343} & {\color{red}1.1390}\\
$wls$ & 0.9948 & 0.9960 & 0.9966 & 0.9956 & 0.9955 & 0.9957 & 0.9955\\
$shr$ & \textbf{0.9899} & \textbf{0.9923} & \textbf{0.9927} & 0.9909 & 0.9912 & \textbf{0.9914} & 0.9910\\
$CCC_H$ & 0.9978 & 0.9981 & 0.9974 & \textbf{0.9905} & \textbf{0.9821} & 0.9946 & \textbf{0.9896}\\
\hline
\end{tabular}
\endgroup
\end{table}

\noindent Rather than considering all the 8 levels of the hierarchy\footnote{The detailed results for all levels, here not presented for space reasons, are available in the on-line appendix.}, we present a more aggregated articulation of the results, by keeping distinct only all, upper, and bottom time series. Even so, it is still possible to say that it is confirmed, when measured by the AvgRelMSE accuracy index as well, the superiority of the $CCC_H$ approach over the optimal combination approaches $wls$ and $shr$, as stated by \cite{Hollyman2021}. However, it should be noted that $CCC_H$ makes use of two different bts base forecasts (automatic ETS and Seasonal Averages), with the risk of an unfair comparison with $wls$ and $shr$: we will come back on this point later.
We add that the overview of the accuracy results does not change when non-negative reconciliation is performed. In fact, the results obtained by using the non-negative reconciliation facilities of the \texttt{R} package \texttt{FoReco} (\citealp{FoReco2021}; see also \citealp{Wickramasuriya2020}, and \citealp{hts2021}), not reported for space reasons but available in the on-line appendix, show that the overall accuracy slightly improves, mainly due to the benefit gained by the forecasts of the many intermittent series at the most disaggregated level, whose unconstrained forecasts often presented negative values. For this reason, and to deal with a more realistic operational context, the extended analysis in the following subsection will be performed by always considering the non-negative variant of the considered forecast reconciliation procedures.


In addition, due to their poor performance, and in order to set a more challenging empirical comparison, in the following we will not consider OLS and TDHP approaches any more, while the $L_1CC$ approach will be only considered as one of the constituent parts of $LCC$ and $CCC$ approaches.

\subsection{Forecast combination based forecast reconciliation using the same bts base forecasts}
From Figure \ref{fig:CCCHscheme}, which simply adapts the $LCC$ and $CCC$ reconciliation scheme in Figure \ref{fig:CCCscheme}, it appears that $CCC_H$ approach is using two different bts base forecasts: in the Upper Level Conditional Coherent reconciliation steps, $\widehat{\bvet}_{SA}$ is used, while in the bottom-up reconciliation $\widehat{\bvet}$ (more precisely, $\widehat{\bvet}_{ETS}$) is used. It must be added that no particular saving of time is obtained by proceeding this way, since ETS base forecasts of all time series (both upper and bottom) are however calculated (and used). Rather, according to $CCC_H$ the base forecasts to be used in each step are \textit{de-facto} appropriately chosen.

In the light of above, we think that if one wishes to exploit both type of bts base forecasts, a sensible starting point 
would be considering the `Seasonal-Averages-based' $LCC$ and $CCC$ reconciled forecasts as well, obtained by simply substituting $\widehat{\bvet}$ with $\widehat{\bvet}_{SA}$ 
in the scheme of Figure \ref{fig:CCCscheme}. Denoting with
$\widetilde{\yvet}^{(l)}_{SA} = \Svet\widetilde{\bvet}^{(l)}_{SA}$, $l=1,\ldots, L$, the $L_lCC$ reconciled forecasts using the seasonal averages as bts base forecasts, and with $\widetilde{\yvet}^{(L+1)}_{SA} = \Svet\widehat{\bvet}_{SA}$ the corresponding bottom-up reconciled forecasts\footnote{When the seasonal averages of the training dataset are used as bts base forecasts, the bottom-up reconciled uts forecasts are equal to the seasonal averages of the uts in the training datasets. Put in other words, unlike the automatic ETS base forecasts, the SA base forecasts of all 525 series are trivially coherent, and need not to be reconciled.}, we obtain the `SA' counterparts of the `ETS-base-forecasts-based' reconciliation approaches described in Figure \ref{fig:CCCscheme}:
\begin{equation}
\label{LCCSA}
\widetilde{\yvet}_{LCC_{SA}}=\displaystyle\frac{1}{L}\displaystyle\sum_{l=1}^{L}\widetilde{\yvet}^{(l)}_{SA} , \qquad
\widetilde{\yvet}_{CCC_{SA}}=\displaystyle\frac{1}{L+1}\displaystyle\sum_{l=1}^{L+1}\widetilde{\yvet}^{(l)}_{SA} .
\end{equation}

\noindent In both cases, the benefit of the forecast combination based forecast reconciliation approaches $LCC$ and $CCC$ is clearly visible: on average across series and forecast horizons, $LCC_{SA}$ reconciled forecasts are always the best performing ones when the seasonal averages are used as bts base forecasts (Table \ref{tab:MSESA}), while when using automatic ETS base forecasts, $LCC_{ETS}$ scores best for all and uts series, while $CCC_{ETS}$ `wins' at the most disaggregated level (Table \ref{tab:MSEETS}).

It must be added that overall the ETS-based $LCC_{ETS}$ and $CCC_{ETS}$ reconciliation perform better than using SA, particularly at the most disaggregated level (Region by PoT, i.e. bts), where the AvgRelMSE's are always less than one, unlike what happens for their SA-based counterparts, which show a stable decrease (AvgRelMSE $>1$) of the forecast accuracy as compared to the benchmark. 

\begin{table}[H]
	\caption{\label{tab:MSESA}\textbf{AvgRelMSE} of $LCC$ and $CCC$ monthly forecast reconciliation
	approaches in the forecasting experiment on the Australian Tourism Demand dataset.
	\textbf{Seasonal averages} of the training sets are used as bts base forecasts. $BU$ identifies the bottom-up approach. Bold entries identify the best performing approaches.
	Red entries identify the approaches worsening the automatic ETS base forecasts' accuracy.}
	\centering
	\begingroup
\footnotesize
	\begin{tabular}[t]{lccccccc}
\hline
		& \multicolumn{7}{c}{Forecast horizon}\\
		Approach & 1 & 2 & 3 & 6 & 12 & 1:6 & 1:12\\
\hline
\multicolumn{8}{c}{\textit{all (525 series)}} \\
$BU$ & {\color{red}1.0341} & {\color{red}1.0345} & {\color{red}1.0323} & {\color{red}1.0249} & {\color{red}1.0029} & {\color{red}1.0295} & {\color{red}1.0205}\\
$L_1CC$ & {\color{red}1.0025} & {\color{red}1.0035} & {\color{red}1.0026} & 0.9962 & 0.9791 & 0.9998 & 0.9930\\
$L_2CC$ & {\color{red}1.0040} & {\color{red}1.0021} & {\color{red}1.0007} & 0.9922 & 0.9797 & 0.9982 & 0.9911\\
$L_3CC$ & {\color{red}1.0145} & {\color{red}1.0136} & {\color{red}1.0132} & {\color{red}1.0072} & 0.9958 & {\color{red}1.0103} & {\color{red}1.0047}\\
$L_4CC$ & {\color{red}1.0275} & {\color{red}1.0296} & {\color{red}1.0277} & {\color{red}1.0259} & {\color{red}1.0121} & {\color{red}1.0270} & {\color{red}1.0227}\\
$L_5CC$ & 0.9931 & 0.9939 & 0.9932 & 0.9890 & 0.9769 & 0.9912 & 0.9865\\
$L_6CC$ & {\color{red}1.0010} & {\color{red}1.0017} & {\color{red}1.0035} & 0.9976 & {\color{red}1.0012} & {\color{red}1.0003} & 0.9986\\
$L_7CC$ & {\color{red}1.0097} & {\color{red}1.0055} & {\color{red}1.0063} & {\color{red}1.0012} & 0.9946 & {\color{red}1.0052} & {\color{red}1.0017}\\
$LCC$ & \textbf{0.9857} & \textbf{0.9865} & \textbf{0.9861} & \textbf{0.9818} & \textbf{0.9736} & \textbf{0.9839} & \textbf{0.9801}\\
$CCC$ & 0.9881 & 0.9889 & 0.9884 & 0.9842 & 0.9753 & 0.9862 & 0.9823\\
\hline
\multicolumn{8}{c}{\textit{upper time series (221 series)}} \\
$BU$ & {\color{red}1.0349} & {\color{red}1.0365} & {\color{red}1.0331} & {\color{red}1.0283} & 0.9980 & {\color{red}1.0311} & {\color{red}1.0213}\\
$L_1CC$ & 0.9761 & 0.9783 & 0.9775 & 0.9749 & 0.9540 & 0.9758 & 0.9699\\
$L_2CC$ & 0.9794 & 0.9761 & 0.9741 & 0.9676 & 0.9561 & 0.9730 & 0.9668\\
$L_3CC$ & 0.9957 & 0.9934 & 0.9934 & 0.9923 & 0.9819 & 0.9917 & 0.9881\\
$L_4CC$ & {\color{red}1.0085} & {\color{red}1.0108} & {\color{red}1.0091} & {\color{red}1.0137} & 0.9988 & {\color{red}1.0104} & {\color{red}1.0085}\\
$L_5CC$ & 0.9599 & 0.9616 & 0.9609 & 0.9621 & 0.9504 & 0.9607 & 0.9584\\
$L_6CC$ & 0.9742 & 0.9748 & 0.9782 & 0.9754 & 0.9928 & 0.9756 & 0.9783\\
$L_7CC$ & 0.9930 & 0.9857 & 0.9878 & 0.9865 & 0.9818 & 0.9884 & 0.9871\\
$LCC$ & \textbf{0.9525} & \textbf{0.9535} & \textbf{0.9535} & \textbf{0.9539} & \textbf{0.9487} & \textbf{0.9527} & \textbf{0.9516}\\
$CCC$ & 0.9564 & 0.9577 & 0.9575 & 0.9581 & 0.9515 & 0.9567 & 0.9554\\
\hline
\multicolumn{8}{c}{\textit{bottom time series (304 series)}} \\
$BU$ & {\color{red}1.0335} & {\color{red}1.0330} & {\color{red}1.0318} & {\color{red}1.0224} & {\color{red}1.0065} & {\color{red}1.0283} & {\color{red}1.0200}\\
$L_1CC$ & {\color{red}1.0221} & {\color{red}1.0223} & {\color{red}1.0212} & {\color{red}1.0120} & 0.9977 & {\color{red}1.0177} & {\color{red}1.0101}\\
$L_2CC$ & {\color{red}1.0223} & {\color{red}1.0214} & {\color{red}1.0205} & {\color{red}1.0105} & 0.9972 & {\color{red}1.0170} & {\color{red}1.0092}\\
$L_3CC$ & {\color{red}1.0283} & {\color{red}1.0286} & {\color{red}1.0279} & {\color{red}1.0181} & {\color{red}1.0060} & {\color{red}1.0240} & {\color{red}1.0169}\\
$L_4CC$ & {\color{red}1.0416} & {\color{red}1.0435} & {\color{red}1.0414} & {\color{red}1.0350} & {\color{red}1.0218} & {\color{red}1.0391} & {\color{red}1.0331}\\
$L_5CC$ & {\color{red}1.0180} & {\color{red}1.0181} & {\color{red}1.0174} & {\color{red}1.0090} & 0.9966 & {\color{red}1.0140} & {\color{red}1.0074}\\
$L_6CC$ & {\color{red}1.0209} & {\color{red}1.0217} & {\color{red}1.0223} & {\color{red}1.0141} & {\color{red}1.0074} & {\color{red}1.0187} & {\color{red}1.0137}\\
$L_7CC$ & {\color{red}1.0219} & {\color{red}1.0202} & {\color{red}1.0200} & {\color{red}1.0120} & {\color{red}1.0040} & {\color{red}1.0176} & {\color{red}1.0125}\\
$LCC$ & \textbf{{\color{red}1.0106}} & \textbf{{\color{red}1.0112}} & \textbf{{\color{red}1.0105}} & \textbf{{\color{red}1.0026}} & \textbf{0.9922} & \textbf{{\color{red}1.0072}} & \textbf{{\color{red}1.0013}}\\
$CCC$ & {\color{red}1.0117} & {\color{red}1.0122} & {\color{red}1.0115} & {\color{red}1.0036} & 0.9930 & {\color{red}1.0083} & {\color{red}1.0023}\\
\hline
	\end{tabular}
	\endgroup
\end{table}

\begin{table}[H]
    \caption{\label{tab:MSEETS}\textbf{AvgRelMSE} of $LCC$ and $CCC$ monthly forecast reconciliation
             	approaches in the forecasting experiment on the Australian Tourism Demand dataset.
             	\textbf{Automatic ETS} are used as bts base forecasts. $BU$ identifies the bottom-up approach. Bold entries identify the best performing approaches.
             	Red entries identify the approaches worsening the automatic ETS base forecasts' accuracy.}
	\centering
	\begingroup
\footnotesize
\setlength\tabcolsep{4pt}
	\begin{tabular}[t]{lccccccc}
\hline
		& \multicolumn{7}{c}{Forecast horizon}\\
		Approach & 1 & 2 & 3 & 6 & 12 & 1:6 & 1:12\\
\hline
\multicolumn{8}{c}{\textit{all (525 series)}} \\
$BU$ & 0.9972 & 0.9940 & 0.9923 & 0.9974 & 1.0008 & 0.9955 & 0.9983\\
$L_1CC$ & 0.9924 & 0.9910 & 0.9902 & 0.9939 & 0.9944 & 0.9922 & 0.9940\\
$L_2CC$ & 0.9920 & 0.9896 & 0.9885 & 0.9890 & 0.9934 & 0.9901 & 0.9912\\
$L_3CC$ & 0.9973 & 0.9950 & 0.9951 & 0.9958 & 0.9944 & 0.9954 & 0.9960\\
$L_4CC$ & {\color{red}1.0107} & {\color{red}1.0113} & {\color{red}1.0112} & {\color{red}1.0103} & {\color{red}1.0048} & {\color{red}1.0109} & {\color{red}1.0095}\\
$L_5CC$ & 0.9876 & 0.9877 & 0.9866 & 0.9908 & 0.9917 & 0.9886 & 0.9907\\
$L_6CC$ & 0.9957 & 0.9959 & 0.9984 & 0.9996 & {\color{red}1.0084} & 0.9980 & {\color{red}1.0004}\\
$L_7CC$ & 0.9994 & 0.9951 & 0.9971 & 0.9953 & 0.9895 & 0.9967 & 0.9950\\
$LCC$ & \textbf{0.9780} & \textbf{0.9772} & \textbf{0.9771} & \textbf{0.9785} & \textbf{0.9791} & \textbf{0.9777} & \textbf{0.9786}\\
$CCC$ & 0.9792 & 0.9781 & 0.9778 & 0.9797 & 0.9808 & 0.9787 & 0.9799\\
\multicolumn{8}{c}{\textit{upper time series (221 series)}} \\
$BU$ & 0.9937 & 0.9861 & 0.9819 & 0.9941 & {\color{red}1.0021} & 0.9894 & 0.9963\\
$L_1CC$ & 0.9823 & 0.9802 & 0.9783 & 0.9880 & 0.9911 & 0.9831 & 0.9883\\
$L_2CC$ & 0.9834 & 0.9787 & 0.9761 & 0.9794 & 0.9902 & 0.9802 & 0.9840\\
$L_3CC$ & 0.9900 & 0.9848 & 0.9848 & 0.9894 & 0.9912 & 0.9869 & 0.9897\\
$L_4CC$ & {\color{red}1.0010} & {\color{red}1.0012} & {\color{red}1.0003} & {\color{red}1.0033} & 0.9992 & {\color{red}1.0019} & {\color{red}1.0025}\\
$L_5CC$ & 0.9738 & 0.9733 & 0.9707 & 0.9816 & 0.9859 & 0.9759 & 0.9817\\
$L_6CC$ & 0.9864 & 0.9860 & 0.9888 & 0.9930 & {\color{red}1.0145} & 0.9895 & 0.9960\\
$L_7CC$ & 0.9947 & 0.9867 & 0.9893 & 0.9914 & 0.9866 & 0.9911 & 0.9912\\
$LCC$ & \textbf{0.9605} & \textbf{0.9585} & \textbf{0.9579} & \textbf{0.9633} & \textbf{0.9684} & \textbf{0.9604} & \textbf{0.9640}\\
$CCC$ & 0.9626 & 0.9599 & 0.9590 & 0.9652 & 0.9708 & 0.9620 & 0.9661\\
\multicolumn{8}{c}{\textit{bottom time series (304 series)}} \\
$BU$ & 0.9998 & 0.9998 & 0.9998 & 0.9999 & 0.9998 & 0.9998 & 0.9998\\
$L_1CC$ & 0.9998 & 0.9989 & 0.9989 & 0.9983 & 0.9967 & 0.9988 & 0.9981\\
$L_2CC$ & 0.9983 & 0.9975 & 0.9976 & 0.9961 & 0.9957 & 0.9973 & 0.9965\\
$L_3CC$ & {\color{red}1.0026} & {\color{red}1.0025} & {\color{red}1.0026} & {\color{red}1.0004} & 0.9968 & {\color{red}1.0016} & {\color{red}1.0005}\\
$L_4CC$ & {\color{red}1.0178} & {\color{red}1.0187} & {\color{red}1.0192} & {\color{red}1.0154} & {\color{red}1.0089} & {\color{red}1.0175} & {\color{red}1.0146}\\
$L_5CC$ & 0.9978 & 0.9983 & 0.9983 & 0.9976 & 0.9960 & 0.9979 & 0.9973\\
$L_6CC$ & {\color{red}1.0026} & {\color{red}1.0032} & {\color{red}1.0055} & {\color{red}1.0045} & {\color{red}1.0040} & {\color{red}1.0043} & {\color{red}1.0037}\\
$L_7CC$ & {\color{red}1.0028} & {\color{red}1.0012} & {\color{red}1.0029} & 0.9982 & 0.9916 & {\color{red}1.0008} & 0.9977\\
$LCC$ & \textbf{0.9910} & \textbf{0.9910} & \textbf{0.9913} & \textbf{0.9898} & \textbf{0.9871} & \textbf{0.9905} & \textbf{0.9893}\\
$CCC$ & 0.9915 & 0.9914 & 0.9917 & 0.9904 & 0.9881 & 0.9910 & 0.9900\\
\hline
	\end{tabular}
	\endgroup
\end{table}

To conclude on this point, we think that a `fair' comparison of the $LCC_{SA} / LCC_{ETS}$ and $CCC_{SA} / CCC_{ETS}$ approaches with the optimal combination forecast reconciliation procedures $wls$ and $shr$, may be established by considering the latter procedures when using either `SA' or `ETS' bts base forecasts. We call these specifications $wls_{SA} / wls_{ETS}$, and $shr_{SA} / shr_{ETS}$, respectively, which will be considered in the next sub-section.


\subsection{Forecast averaging vs. pooling of reconciled forecasts}
A simple, natural way to exploit the features of different bts base forecasts consists in considering a reconciled vector of forecasts obtained by averaging each approach using the bts base forecasts from both models (SA and ETS). This means to consider the vector of averaged forecasts
\begin{equation}
	\label{avrecfor}
\widetilde{\yvet}_{\overline{{\rm rec}}} = \displaystyle\frac{\widetilde{\yvet}_{{\rm rec}_{SA}} + \widetilde{\yvet}_{{\rm rec}_{ETS}}}{2} , \quad
{\rm rec} = wls, shr, CCC, LCC.
\end{equation}

\noindent Coherently with this framework, the $CCC_H$ approach can be seen as the forecast combination of the $2(L+1)$ forecasts $\widetilde{\yvet}^{(l)}_{ETS}$ and $\widetilde{\yvet}^{(l)}_{SA}$, $l=1,\ldots, L+1$, where the $L$ forecasts' vectors $\widetilde{\yvet}^{(l)}_{ETS}$, $l=1, \ldots, L$, and the single forecast vector $\widetilde{\yvet}^{L+1}_{SA}$, are given zero weights, while the remaining forecasts are equally weighted by $\displaystyle\frac{1}{L+1}$.

This choice may be seen as a forecast pooling (\citealp{Hendry2004}, \citealp{Marcellino2004}, \citealp{Geweke2011}), a forecast combination procedure according to which from the complete set of forecasts, only a subset is deemed relevant to be combined. \cite{KourentzesBarrowPetropoulos2019} investigate pooling for business forecasting, claiming ``that the forecast selection criteria and the different approaches that are used to combine forecasts, can be considered as two independent types of operations, that follow pooling. (...) Forecast selection and forecast combinations can be seen as two extremes of a spectrum that is defined by forecast pooling, combined with some selection/weighting operator''.\footnote{An interesting Machine Learning approach to model selection in hierarchical forecasting has been recently proposed by \cite{Abolghasemi2020}.}
\cite{Aiolfi2006} state that pooling can be beneficial, but recognise that the methods proposed depend on multiple subjective choices by the modeller.
We think that this is just the case of the $CCC_H$ approach, which is grounded on a forecast pooling whose motivation might appear somehow subjective, therefore not immediately generalizable to different forecasting situations. Rather, simple complete combinations like either $\overline{CCC}$ or $\overline{LCC}$ seem to be very simple as well, able to exploit the diversity of forecasts coming from different models, and less prone to subjective choices. In addition, the comparison with the forecasting accuracy of $\overline{wls}$ and $\overline{shr}$ approaches seems to be logically well founded. 

Table \ref{tab:AvgRelMSE} sets out the AvgRelMSE for the reconciliation approaches described so far, distinct by all, upper and bottom variables, and grouped in such a way to highlight the common background in terms of the bts base forecasts used. 
For completeness, in Table \ref{tab:AvgRelMSE} we consider the forecast accuracy (relative to the ETS base forecasts) of the seasonal averages as well, as these simple forecasts are at the same time cross-sectionally coherent, and part of the `SA-based' forecast reconciliation approaches.

In summary, we find that:
\begin{itemize}
\item Forecast averaging of the $LCC$ reconciled forecasts (i.e., $\overline{LCC}$) gives the best results for most forecast horizons: in 7 cases out 7 across all 525 series, 6 out 7 across the 221 upper series, and 2 out 7 across the 304 bottom series. In this last case,  $\overline{CCC}$ ranks first, the differences with $\overline{LCC}$ being at the fourth decimal point.
\item Forecast averaging of the optimal combination forecasts approaches  ($\overline{shr}$ and $\overline{shr}$) always performs better than $CCC_H$ at the most disaggregate levels (304 bts series), while $CCC_H$ scores better than $\overline{wls}$ and $\overline{shr}$ across the 221 more aggregated upper time series.
\item When bts base forecasts from a single model (either SA or ETS) are used, $shr$ performs on average best across all series. This result comes from a clear superiority for the most aggregated series in the upper levels of the hierarchy, while at the bottom level the $shr$ and $LCC$ AvgRelMSE's look very similar.
\end{itemize}


\begin{table}[H]
	\caption{\label{tab:AvgRelMSE}\textbf{AvgRelMSE} of monthly reconciled forecasts in the forecasting experiment on the Australian Tourism Demand dataset. Optimal combination, $LCC$, and $CCC$ reconciliation approaches, using seasonal averages and/or automatic ETS as bts base forecasts. Bold entries identify the best performing approaches independently of the base forecasts used. Italic entries identify the best performing approach using the same base forecasts (either SA or automatic ETS). Red entries identify the approaches worsening the automatic ETS base forecasts' accuracy.}
	\centering
	\begingroup
\footnotesize
\def\arraystretch{0.92}
	\begin{tabular}[t]{cccccccccc}
\hline
 &\multicolumn{2}{c}{Base forecasts$^*$} & \multicolumn{7}{c}{Forecast horizon}\\
Approach$^{**}$ & uts & bts & 1 & 2 & 3 & 6 & 12 & 1:6 & 1:12\\
\hline
\multicolumn{10}{c}{\textit{all (525 series)}} \\
SA               & SA   & SA      & {\color{red} 1.0341} & {\color{red} 1.0345} & {\color{red} 1.0323} & {\color{red} 1.0249} & {\color{red} 1.0029} & {\color{red} 1.0295} & {\color{red} 1.0205}\\
\hline
$wls_{SA}$            & ETS  & SA      & 0.9892 & 0.9904 & 0.9902 & 0.9867 & 0.9809 & 0.9883 & 0.9853\\
$shr_{SA}$            & ETS  & SA      & \textit{0.9833} & \textit{0.9855} & \textit{0.9851} & \textit{0.9812} & 0.9754 & \textit{0.9831} & \textit{0.9800}\\
$CCC_{SA}$            & ETS  & SA      & 0.9881 & 0.9889 & 0.9884 & 0.9842 & 0.9753 & 0.9862 & 0.9823\\
$LCC_{SA}$            & ETS  & SA      & 0.9857 & 0.9865 & 0.9861 & 0.9818 & \textit{0.9736} & 0.9839 & 0.9801\\
\hline
$wls_{ETS}$            & ETS & ETS     & 0.9801 & 0.9801 & 0.9802 & 0.9806 & 0.9816 & 0.9802 & 0.9806\\
$shr_{ETS}$            & ETS & ETS     & \textit{0.9739} & \textit{0.9757} & \textit{0.9754} & \textit{0.9748} & \textit{0.9768} & \textit{0.9750} & \textit{0.9752}\\
$CCC_{ETS}$            & ETS & ETS     & 0.9792 & 0.9781 & 0.9778 & 0.9797 & 0.9808 & 0.9787 & 0.9799\\
$LCC_{ETS}$            & ETS & ETS     & 0.9780 & 0.9772 & 0.9771 & 0.9785 & 0.9791 & 0.9777 & 0.9786\\
\hline
$\overline{wls}$ & ETS & SA \& ETS & 0.9715 & 0.9719 & 0.9719 & 0.9703 & 0.9685 & 0.9709 & 0.9697\\
$\overline{shr}$ & ETS & SA \& ETS & 0.9698 & 0.9716 & 0.9711 & 0.9686 & 0.9671 & 0.9699 & 0.9684\\
$CCC_H$          & ETS & SA \& ETS & 0.9757 & 0.9759 & 0.9753 & 0.9721 & 0.9659 & 0.9737 & 0.9708\\
$\overline{CCC}$ & ETS & SA \& ETS & 0.9623 & 0.9622 & 0.9617 & 0.9608 & 0.9583 & 0.9612 & 0.9602\\
$\overline{LCC}$ & ETS & SA \& ETS & \textbf{0.9618} & \textbf{0.9618} & \textbf{0.9615} & \textbf{0.9602} & \textbf{0.9577} & \textbf{0.9608} & \textbf{0.9596}\\
\hline
\multicolumn{10}{c}{\textit{upper time series (221 series)}} \\
SA               & SA   & SA      & {\color{red} 1.0349} & {\color{red} 1.0365} & {\color{red} 1.0331} & {\color{red} 1.0283} & 0.9980 & {\color{red} 1.0311} & {\color{red} 1.0213}\\
\hline
$wls_{SA}$            & ETS             & SA     & 0.9584 & 0.9603 & 0.9608 & 0.9623 & 0.9603 & 0.9602 & 0.9602\\
$shr_{SA}$            & ETS             & SA     & \textit{0.9485} & \textit{0.9516} & \textit{0.9512} & \textit{0.9514} & 0.9498 & \textit{0.9505} & \textit{0.9499}\\
$CCC_{SA}$            & ETS      & SA     & 0.9564 & 0.9577 & 0.9575 & 0.9581 & 0.9515 & 0.9567 & 0.9554\\
$LCC_{SA}$            & ETS      & SA     & 0.9525 & 0.9535 & 0.9535 & 0.9539 & \textit{0.9487} & 0.9527 & 0.9516\\
\hline
$wls_{ETS}$            & ETS            & ETS    & 0.9612 & 0.9595 & 0.9595 & 0.9629 & 0.9687 & 0.9610 & 0.9636\\
$shr_{ETS}$            & ETS            & ETS    & \textit{0.9532} & \textit{0.9541} & \textit{0.9533} & \textit{0.9554} & \textit{0.9626} & \textit{0.9544} & \textit{0.9566}\\
$CCC_{ETS}$            & ETS     & ETS    & 0.9626 & 0.9599 & 0.9590 & 0.9652 & 0.9708 & 0.9620 & 0.9661\\
$LCC_{ETS}$            & ETS     & ETS    & 0.9605 & 0.9585 & 0.9579 & 0.9633 & 0.9684 & 0.9604 & 0.9640\\
\hline
$\overline{wls}$ & ETS & SA \& ETS & 0.9547 & 0.9548 & 0.9551 & 0.9574 & 0.9594 & 0.9554 & 0.9567\\
$\overline{shr}$ & ETS & SA \& ETS & 0.9474 & 0.9494 & 0.9487 & 0.9496 & 0.9522 & 0.9488 & 0.9494\\
$CCC_H$          & ETS & SA \& ETS & 0.9473 & 0.9473 & 0.9467 & 0.9484 & \textbf{0.9449} & 0.9468 & 0.9466\\
$\overline{CCC}$ & ETS & SA \& ETS & 0.9438 & 0.9433 & 0.9427 & 0.9461 & 0.9466 & 0.9438 & 0.9454\\
$\overline{LCC}$ & ETS & SA \& ETS & \textbf{0.9426} & \textbf{0.9422} & \textbf{0.9419} & \textbf{0.9448} & 0.9455 & \textbf{0.9427} & \textbf{0.9441}\\
\hline
\multicolumn{10}{c}{\textit{bottom time series (304 series)}} \\
SA               & SA   & SA      & {\color{red} 1.0335} & {\color{red} 1.0330} & {\color{red} 1.0318} & {\color{red} 1.0224} & {\color{red} 1.0065} & {\color{red} 1.0283} & {\color{red} 1.0200}\\
\hline
$wls_{SA}$            & ETS      & SA     & {\color{red} 1.0122} & {\color{red} 1.0128} & {\color{red} 1.0121} & {\color{red} 1.0049} & 0.9961 & {\color{red} 1.0093} & {\color{red} 1.0040}\\
$shr_{SA}$            & ETS      & SA     & \textit{{\color{red} 1.0094}} & \textit{{\color{red} 1.0109}} & {\color{red} 1.0105} & {\color{red}1.0034} & 0.9944 & {\color{red} 1.0075} & {\color{red} 1.0024}\\
$CCC_{SA}$            & ETS      & SA    & {\color{red} 1.0117} & {\color{red} 1.0122} & {\color{red} 1.0115} & {\color{red} 1.0036} & 0.9930 & {\color{red} 1.0083} & {\color{red} 1.0023}\\
$LCC_{SA}$            & ETS      & SA     & {\color{red} 1.0106} & {\color{red} 1.0112} & \textit{{\color{red} 1.0105}} & \textit{{\color{red} 1.0026}} & \textit{0.9922} & \textit{{\color{red} 1.0072}} & \textit{{\color{red} 1.0013}}\\
\hline
$wls_{ETS}$            & ETS     & ETS    & 0.9941 & 0.9953 & 0.9956 & 0.9937 & 0.9911 & 0.9945 & 0.9931\\
$shr_{ETS}$            & ETS     & ETS    & \textit{0.9892} & 0.9917 & 0.9917 & \textit{0.9890} & 0.9873 & \textit{0.9903} & 0.9889\\
$CCC_{ETS}$            & ETS     & ETS    & 0.9915 & 0.9914 & 0.9917 & 0.9904 & 0.9881 & 0.9910 & 0.9900\\
$LCC_{ETS}$            & ETS     & ETS    & 0.9910 & \textit{0.9910} & \textit{0.9913} & 0.9898 & \textit{0.9871} & 0.9905 & \textit{0.9893}\\
\hline
$\overline{wls}$ & ETS & SA \& ETS & 0.9839 & 0.9845 & 0.9843 & 0.9797 & 0.9752 & 0.9824 & 0.9793\\
$\overline{shr}$ & ETS & SA \& ETS & 0.9865 & 0.9880 & 0.9878 & 0.9827 & 0.9780 & 0.9856 & 0.9824\\
$CCC_H$          & ETS & SA \& ETS & 0.9969 & 0.9972 & 0.9965 & 0.9897 & 0.9814 & 0.9938 & 0.9887\\
$\overline{CCC}$ & ETS & SA \& ETS & \textbf{0.9759} & \textbf{0.9761} & \textbf{0.9757} & 0.9717 & 0.9669 & \textbf{0.9740} & 0.9711\\
$\overline{LCC}$ & ETS & SA \& ETS & 0.9761 & 0.9762 & 0.9759 & \textbf{0.9716} & \textbf{0.9667} & 0.9741 & \textbf{0.9710}\\
\hline
\multicolumn{10}{l}{\scriptsize $^*$ SA: seasonal averages; ETS: automatic ETS forecasts.} \\[-.1cm]
\multicolumn{10}{l}{\scriptsize $^{**}$ $CCC_H$: base forecasts as in Hollyman et al. (2021).}
	\end{tabular}
	\endgroup
\end{table}





\begin{figure}[H]
	\centering
	\begin{subfigure}{1\textwidth}
		\centering
		\includegraphics[width=0.6\textwidth]{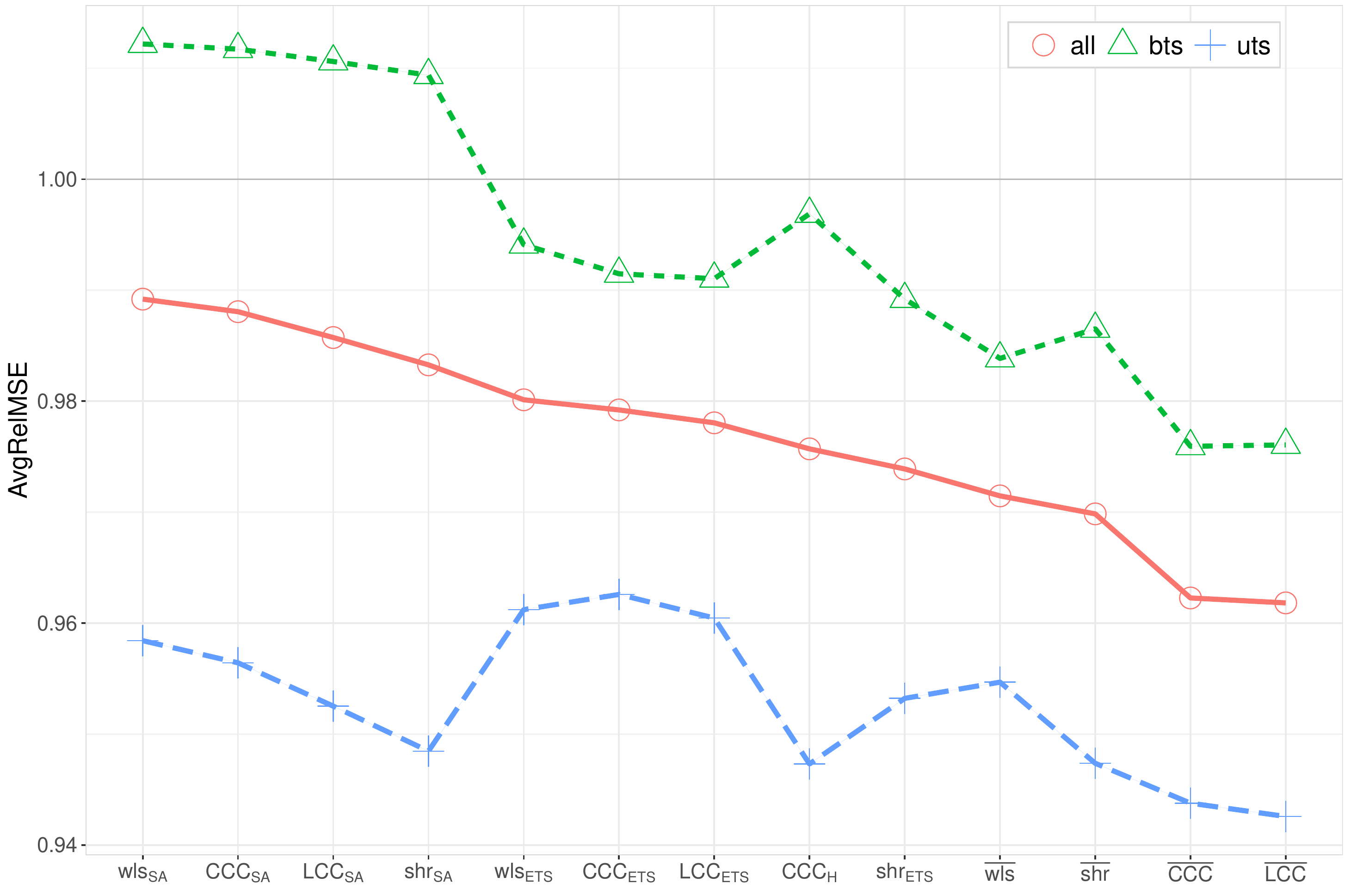}
		\caption{forecast horizon $h=$ 1}
		\label{fig:FinalFigMSEh1}
	\end{subfigure}

\bigskip\bigskip

	\begin{subfigure}{1\textwidth}
		\centering
		\includegraphics[width=0.6\textwidth]{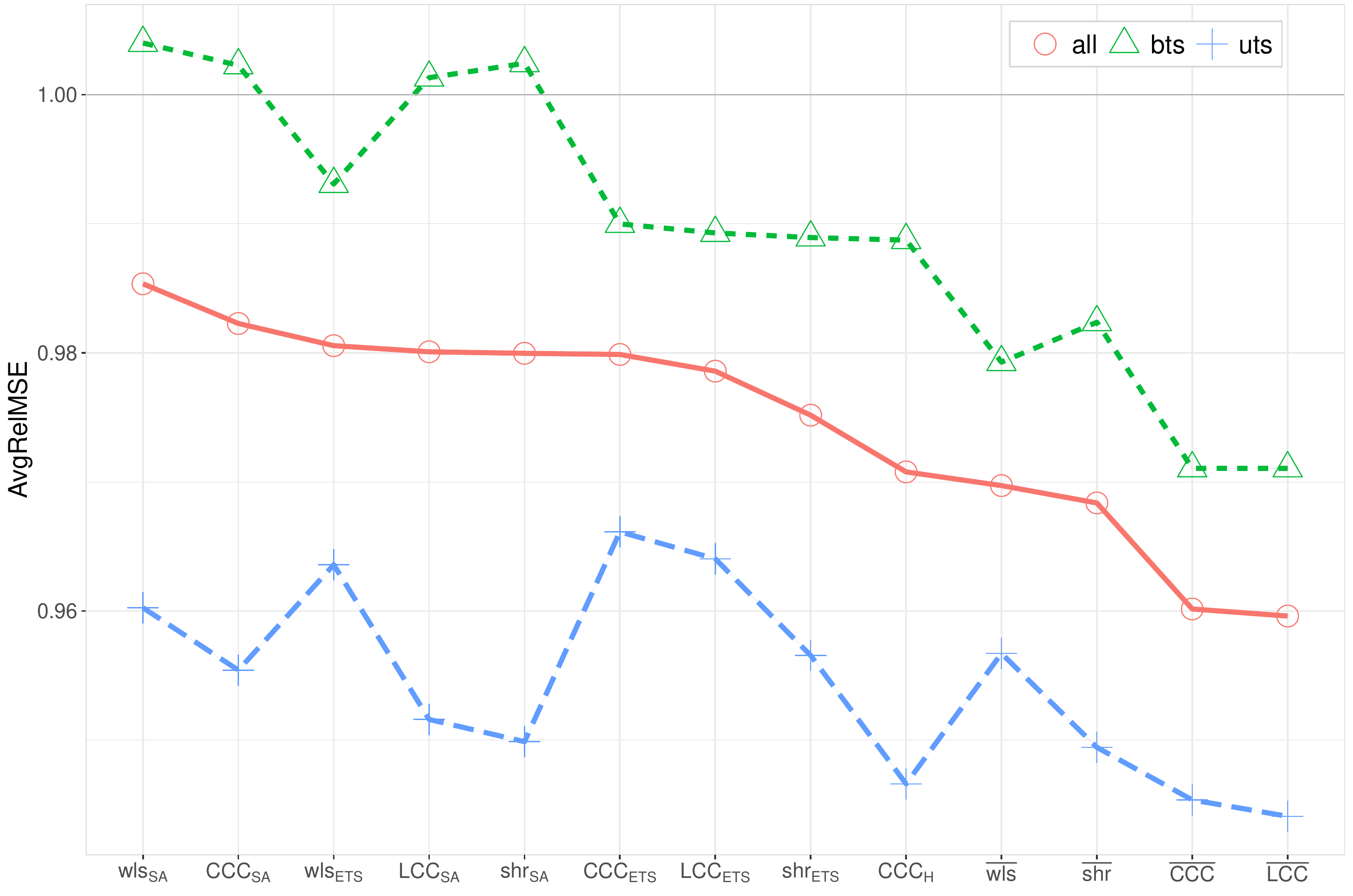}
		\caption{forecast horizon $h=$ 1:12}
		\label{fig:FinalFigMSEh1-12}
	\end{subfigure}
	\caption{\textbf{AvgRelMSE} of Optimal combination, $LCC$, and $CCC$ reconciliation approaches, using seasonal averages and/or automatic ETS as bottom time series base forecasts. (a) forecast horizon $h=$ 1, (b) forecast horizon $h=$ 1:12 (values from the second and last column of Table \ref{tab:AvgRelMSE}, respectively)}
	\label{fig:FinalFigMSE}
\end{figure}

Figure \ref{fig:FinalFigMSE} gives an interesting visual summary of the results for two forecast horizons: $h=$ 1 and $h=$ 1:12. The `SA-based' reconciliation approaches are located to the left part of the graphs, due to the low quality of the bts base forecasts. They are followed by the `ETS-based' reconciled forecasts, which benefit from the higher quality of the bts base forecasts. The $CCC_H$ approach clearly separates the single-model bts base forecasts from the multiple (two, in this application) base forecasts combination based reconciliation, where the forecast combination device of $LCC$ and $CCC$ approaches gives better results than the average of a couple of optimal combination reconciled forecasts.

\begin{figure}[H]
	\centering
	\begin{subfigure}{1\textwidth}
		\centering
		\includegraphics[width=0.85\textwidth]{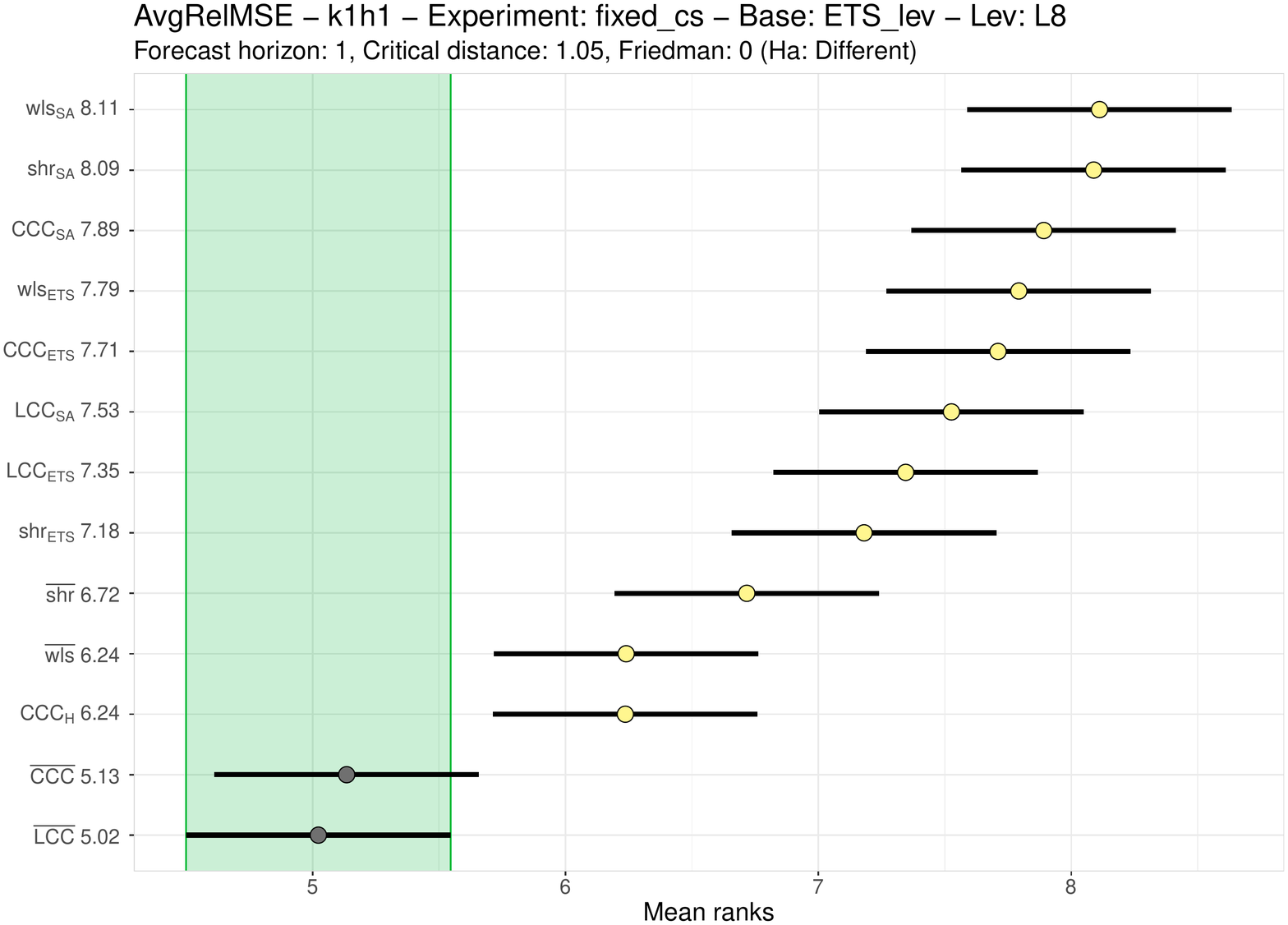}
		\caption{forecast horizon $h=$ 1}
		\label{fig:MSEmcb1_bts}
	\end{subfigure}

\bigskip

	\begin{subfigure}{1\textwidth}
		\centering
		\includegraphics[width=0.85\textwidth]{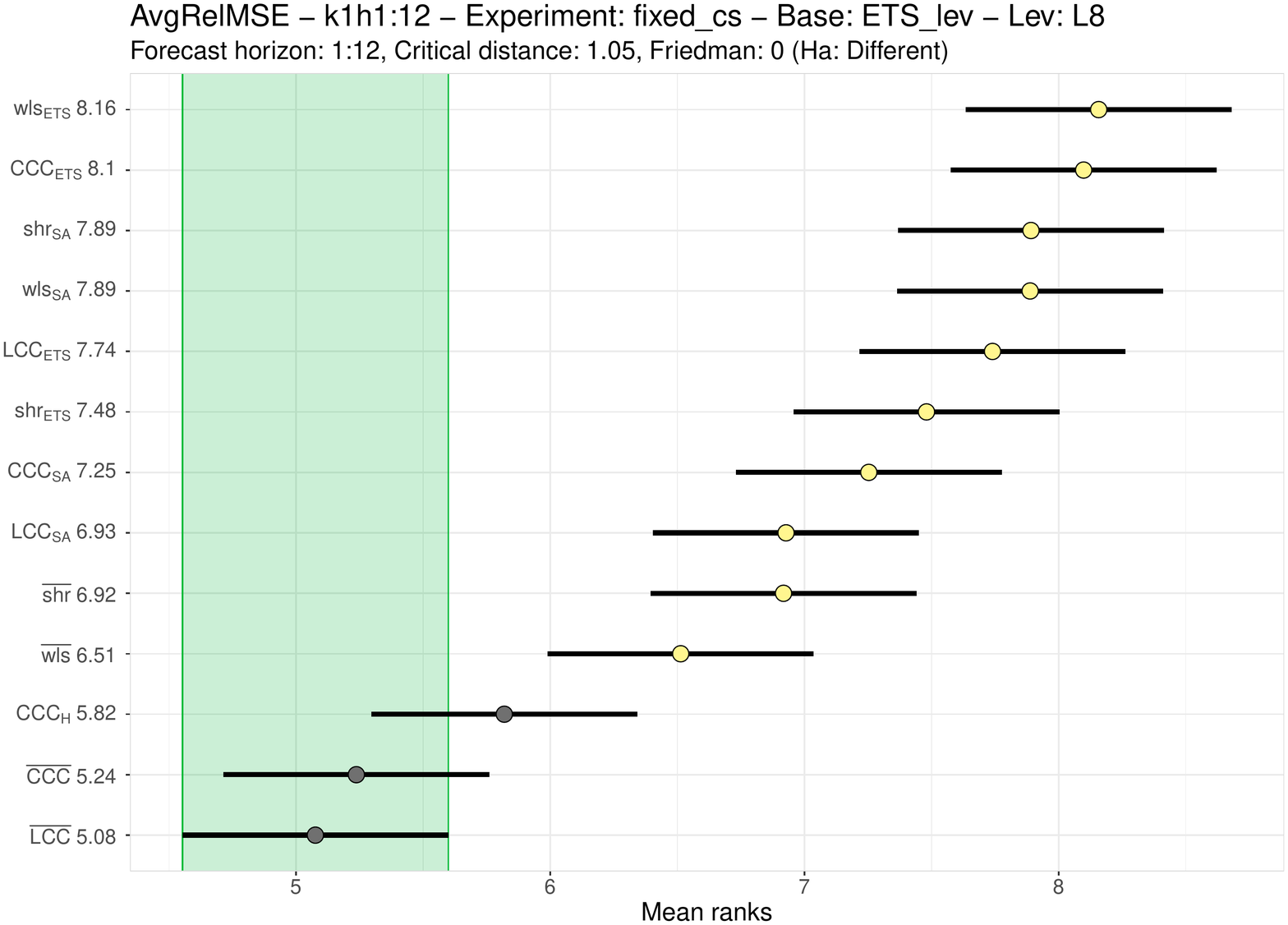}
		\caption{forecast horizon $h=$ 1:12}
		\label{fig:MSEmcb1-12_bts}
	\end{subfigure}
	\caption{MCB Nemenyi test results: average ranks and 95\% confidence intervals for the 304 bts forecasts. The reconciliation approaches are sorted vertically according to the \textbf{MSE} mean rank for forecast horizon $h=$ 1 (a), and forecast horizon $h=$ 1:12 (b). The mean rank of each approach is displayed to the right of their names. If the intervals of two forecast reconciliation procedures do not overlap, this indicates a statistically different performance. Thus, approaches that do not overlap with the green interval are considered significantly worse than the best, and vice-versa.
	}
	\label{fig:MSEmcb}
\end{figure}

Furthermore, it is confirmed and somehow reinforced the observation of \cite{Hollyman2021} that the gains from combination are particularly valuable when existing forecasts are of poor quality. Figure \ref{fig:MSEmcb} shows the Multiple Comparison with the Best (MCB) Nemenyi test for the 304 bts, which form the noisiest section of the hierarchy. After that the Friedman test has shown that the considered forecasting approaches are different, $\overline{LCC}$ and $\overline{CCC}$ look (better than, and) significantly different from the other reconciliation forecasts at forecast horizon $h=$ 1. These approaches still rank best for $h=$ 1:12, with $CCC_H$ forecasts resulting not significantly different from them.




\section{Conclusions}
\label{sec: conclusions}

In this paper, the $LCC$ cross-sectional forecast reconciliation approach recently proposed by \cite{Hollyman2021} has been re-visited. It was shown that it can be interpreted as the solution to a linearly constrained quadratic minimization problem with exogenous constraints, given by the upper time series base forecasts. We also provide the expressions valid for a level conditional coherent reconciliation with endogenous constraints, where the upper time series forecasts are no longer considered as binding constraints, but are admitted to be revised in view of their variability.

The forecasting experiment on the Australian Tourism Demand dataset by \cite{Hollyman2021} has been extended accordingly, in order to compare the accuracy performance of the state-of-the-art optimal combination forecast reconciliation procedures (\citealp{Wickramasuriya2019}) with those offered by simple averaging of the $LCC$ reconciled forecasts. Results have been found (i) considering the non-negativity issues posed by the data, (ii) using the relative accuracy metrics recommended by \cite{Davydenko2013}, and (iii) taking into account the role played by the bts base forecasts in the $LCC$ and $CCC$ forecast combination, which allows to establish a `fair' comparison with the optimal combination forecasts reconciliation procedure.

The learned lesson is that pooling reconciled forecasts plays a positive and important role in forecast reconciliation also: applying simple pooling techniques, while fulfilling all the cross-sectional constraints, improves the quality of the single constituent forecasts (\citealp{Abouarghoub2018}), in agreement with the vast amount of empirical evidence of the last five decades in the field of forecast combination (\citealp{Bates1969}, \citealp{Clemen1989}, \citealp{Timmermann2006}).

Furthermore, the original intuition by \cite{Hollyman2021} of combining bottom time series base forecasts from different models (e.g., seasonal averages and automatic ETS) has been conveniently re-stated (and somehow re-inforced) by considering a forecast averaging strategy involving all the available forecasts, which resulted in $LCC$ and $CCC$ reconciled forecasts better performing than those produced by the optimal combination reconciliation approaches, and the forecast pooling strategy of \cite{Hollyman2021}.

In this paper, only simple forecast averaging has been considered, in line with the idea that it is generally not worse (and often it is better) than more sophisticated weighting schemes (\citealp{Genre2013}).
Nevertheless, we think it would be interesting, and potentially fruitful, to consider alternative forecast pooling methods (\citealp{KourentzesBarrowPetropoulos2019}, \citealp{Lichtendahl2020}).
In addition, in order to exploit information differences and mitigate model uncertainty, the $LCC$ approach could be easily extended to combine forecasts from multiple temporal aggregation levels (\citealp{Athanasopoulos2017}), provided a sensible forecast error covariance matrix be considered (\citealp{Nystrup2020, Nystrup2021}). By continuing on this path,
since leveraging both cross-sectional and temporal hierarchies using cross-temporal reconciliation approaches has shown to be effective in order to forecast a linearly constrained multiple time series (\citealp{Kourentzes2019}, \citealp{DiFonzoGiro2020}), forecast combination based forecast reconciliation could be adapted to this challenging framework as well, in order to gain predictive accuracy.
Finally, as in practical applications a thorough forecast accuracy evaluation need to deal with predictive distibutions rather than point forecasts, the forecast combination approach to probabilistic forecast reconciliation (\citealp{Jeon2019}, \citealp{Panagiotelis2020b}, \citealp{Yang2020a, Yang2020b}, \citealp{Wickramasuriya2021a}) is a valuable topic worth considering for future research.




\section*{Appendix 1. Level Conditional Coherent forecast reconciliation: the original formulation by Hollyman et al. (2021)}

\subsection*{Level-1 Conditional Coherent forecast reconciliation}
For $l=1$, in order to transform the bts base forecasts $\widehat{\bvet}$ in reconciled forecasts $\widetilde{\bvet}^{(1)}$ conditional to $\widehat{y}_1$, \cite{Hollyman2021} consider the $\left[(n_b+1) \times (n_b+1)\right]$ matrix $\Avet_1$:
\[
\Avet_1 = \begin{bmatrix}
	0 & {\bf 1}'_{n_b} \\
	\pvet & \Ivet_{n_b}
\end{bmatrix} ,
\]
where $\pvet$ is a $(n_b \times 1)$ vector of \textit{combination weights} $p_i$, $i=1,\ldots,n_b$, $0 < p_i < 1$, $\displaystyle\sum_{i=1}^{n_b} p_i = 1$.
As it is immediately recognized, such a matrix may be obtained by `augmenting' the level-1 structural summation matrix $\Svet_1$, putting the vector $\left[0 \; \pvet'\right]'$ on its left side.
Let $\bar{\Gvet}_1$ be a $\left[(n_b+1) \times (n_b+1)\right]$ matrix linked to $\Avet_1$ by the relationship
\[
\bar{\Gvet}_1 \Avet_1 = \Ivet_{(n_b+1)} .
\]
By solving the previous relationship wrt $\bar{\Gvet}_1$, i.e. $\bar{\Gvet}_1 = \Avet_1^{-1}$, it is possible to get the weights to be used to combine the base forecasts of the bts, and the $n_b$ lower rows of matrix $\bar{\Gvet}_1$ compose the $\left[n_b \times (n_b +1)\right]$ matrix $\Gvet_1$, transforming the bts base forecasts in coherent forecasts.
It can be easily checked that
$$
\bar{\Gvet}_1 = \Avet_1^{-1} = \begin{bmatrix}
	-1 & {\bf 1}'_{n_b} \\\pvet & \left(\Ivet_{n_b} - \pvet{\bf 1}'_{n_b}\right)
\end{bmatrix} ,
$$
and thus we can write $\widetilde{\bvet}^{(1)} = \Gvet_1\widehat{\yvet}_1$, where
$
\Gvet_1 = \left[\pvet \;\; \left(\Ivet_{n_b} - \pvet{\bf 1}'_{n_b}\right)\right] 
$
is a
$\left[n_b \times (n_b + 1)\right]$ matrix obtained by removing the first row of $\bar{\Gvet}_1$.

All the reconciled forecasts are thus given by $\widetilde{\yvet}^{(1)} = \Svet\widetilde{\bvet}^{(1)} = \Svet\Gvet_1\widehat{\yvet}_1$. In order to express the vector $\widetilde{\yvet}^{(1)}$ as a transformation of all the base forecasts, not only the ones for the top-level series and the bts, $n_a - 1$ zero columns have to be inserted between the first and the second column of matrix $\Gvet_1$, thus obtaining the $(n_b \times n)$ matrix $\Gvet^{(1)}$, such that:
$$
\widetilde{\yvet}^{(1)} = \Gvet^{(1)}\widehat{\yvet} .
$$
For the simple hierarchical time series considered as an example in Figure \ref{Fig:hts1}, we would have the $(5 \times 8)$ matrix
\[
\Gvet^{(1)} = \left[\pvet \;\; \Zerovet_{5 \times 2} \; \; \left(\Ivet_{5} - \pvet{\bf 1}'_{5}\right)\right] .
\]
It is worth noting that, since $\Gvet^{(1)}\Svet = \Ivet_5$, this matrix satisfies the unbiasedness condition for the reconciled forecasts (Athanasopoulos et al., 2009).


\subsection*{Extension for $l>1$}
Let's focus now on a generic level $l$, $l=1,\ldots L$, of the hierarchical/grouped time series, and denote with $\widetilde{\yvet}_l$ the $\left[(n_l+n_b) \times 1\right]$ vector containing the reconciled forecasts conditional to the level-$l$ uts base forecasts. In other words, the uts reconciled forecasts in $\widetilde{\yvet}_l$ are equal to the base forecasts of the corresponding uts in $\widehat{\yvet}_l$ (i.e., $\widetilde{\avet}_l = \widehat{\avet}_l$). Then, denote $\widetilde{\bvet}^{(l)}$, $l=1,\ldots, L$, the bts reconciled forecasts conditional to the base forecasts of the level-$l$ series.

In analogy to the case $l=1$, \cite{Hollyman2021} propose to compute
\[
\widetilde{\bvet}^{(l)} = \Gvet_l\widehat{\yvet}_l ,
\]
where $\Gvet_l$ is a $\left[n_b \times (n_l+n_b)\right]$ matrix built as follows.

Let $\Pvet_l$ be the $(n_b \times n_l)$ matrix (\ref{Plmat}) containing the weights of each forecasts in the $n_l$ elementary hierarchies linking each of the $n_l$ level-$l$ series to their `afferent' bts, and define
the $\left[(n_l + n_b) \times (n_l + n_b)\right]$ matrix $\Avet_l$ as:
\begin{equation*}
	\Avet_l =  
	\begin{bmatrix}
		\Zerovet_{n_l \times n_l} & \Cvet_l \\
		\Pvet_l & \Ivet_{n_b}
	\end{bmatrix} ,
\end{equation*}
where $\Cvet_l$ is the cross-sectional (contemporaneous) aggregation matrix, with dimension $(n_l \times n_b)$, mapping the $n_b$ bts into the $n_l$ level-$l$ aggregated series, such that $\Cvet_l\Pvet_l = \Ivet_{n_l}$.

$\Avet_l$ is a $2 \times 2$ block-partitioned matrix, with a null upper-left block, whose inverse is given by (\citealp{Lu2002}):
\[
\Avet_l^{-1} = \begin{bmatrix}
	-\left(\Cvet_l\Pvet_l\right)^{-1} & \left(\Cvet_l\Pvet_l\right)^{-1}\Cvet_l \\
	\Pvet_l\left(\Cvet_l\Pvet_l\right)^{-1} & \left[\Ivet_{n_b} - \Pvet_l\left(\Cvet_l\Pvet_l\right)^{-1}\Cvet_l\right]
\end{bmatrix} = 
\begin{bmatrix}
	-\Ivet_{n_l} & \Cvet_l \\
	\Pvet_l & \left(\Ivet_{n_b} - \Pvet_l\Cvet_l\right)
\end{bmatrix},
\]
a very simple expression to compute, which does not need any matrix inversion.

The wished transformation matrix $\Gvet_l$, which has dimension $\left[n_b \times (n_b+n_l)\right]$, allowing to calculate bts reconciled forecasts in line with the $n_l$ base forecasts of the level-$l$ series, can be obtained by simply discarding the top $n_l$ rows of matrix $\Avet_l^{-1}$:
$\Gvet_l = \begin{bmatrix}
	\Pvet_l & \left(\Ivet_{n_b} - \Pvet_l\Cvet_l\right)
\end{bmatrix}$.
The \emph{level-$l$ conditional bts reconciled forecasts} are thus given by:
\begin{equation}
	\label{levlrec}
	\begin{array}{rcl}
		\widetilde{\bvet}^{(l)} & = &\Pvet_l\widehat{\avet}_l + \left(\Ivet_{n_b} - \Pvet_l\Cvet_l\right)\widehat{\bvet} \\
		& = & \widehat{\bvet} + \Pvet_l\left(\widehat{\avet}_l - \Cvet_l\widehat{\bvet}\right)
	\end{array}, \quad l = 1, \ldots, L ,
\end{equation}

\noindent and the vector $\widetilde{\yvet}^{(l)}$ of all the reconciled forecasts conditional to the level-$l$ base forecasts is easily obtained as:
\[
\widetilde{\yvet}^{(l)} = \Svet\widetilde{\bvet}^{(l)} , \quad l = 1, \ldots, L .
\]

\section*{Appendix 2. Derivation of $\widetilde{\bvet}^{(l)}_{\text{LCC}}$}
Consider the lagrangean function
\[
L\left(\bvet,\lambdavet\right) = \bvet'\Wvet_b^{-1}\bvet - 2\widehat{\bvet}'\Wvet_b^{-1}\bvet + \widehat{\bvet}'\Wvet_b^{-1}\widehat{\bvet} + 2\lambdavet'\left(\Cvet_l\bvet - \avet_l\right) .
\]
The first order conditions are given by:
\[
\begin{array}{rcrcc}
	\displaystyle\frac{\partial L}{\partial \bvet} & = & \Wvet_b^{-1}\bvet - 2\Wvet_b^{-1}\widehat{\bvet} +
	2\Cvet_l'\lambdavet & = & \Zerovet \\[.5cm]
	\displaystyle\frac{\partial L}{\partial \lambdavet} & = & \Cvet_l\bvet - \avet_l & = & \Zerovet
\end{array} .
\]
After simplification and re-arrangement of the known terms on the right side of the expression, it is:
\[
\begin{array}{lcc}
	\Wvet_b^{-1}\bvet + \Cvet_l'\lambdavet & = & \Wvet_b^{-1}\widehat{\bvet} \\[.5cm]
	\Cvet_l\bvet                           & = & \avet_l
\end{array} \quad \longrightarrow \quad
\begin{bmatrix}
	\Wvet_b^{-1} & \Cvet_l' \\
	\Cvet_l & \Zerovet
\end{bmatrix}
\begin{bmatrix}
\bvet \\ \lambdavet
\end{bmatrix} =
\begin{bmatrix}
\Wvet_b^{-1}\widehat{\bvet} \\ \avet_l
\end{bmatrix} .
\]
The solution to the system is thud given by
\[
\begin{bmatrix}
	\bvet \\ \lambdavet
\end{bmatrix} =
\begin{bmatrix}
	\Wvet_b^{-1} & \Cvet_l' \\
	\Cvet_l & \Zerovet
\end{bmatrix}^{-1}
\begin{bmatrix}
	\Wvet_b^{-1}\widehat{\bvet} \\ \avet_l
\end{bmatrix} .
\]
The inverse matrix is given by (\citealp{Lu2002}):
\[
\begin{bmatrix}
	\Wvet_b^{-1} & \Cvet_l' \\
	\Cvet_l & \Zerovet
\end{bmatrix}^{-1} =
\begin{bmatrix}
\left(\Wvet_b - \Wvet_b\Cvet_l'\left(\Cvet_l\Wvet_b\Cvet_l'\right)^{-1}\Cvet_l\Wvet_b\right) & \Wvet_b\Cvet_l'\left(\Cvet_l\Wvet_b\Cvet_l'\right)^{-1} \\
\left(\Cvet_l\Wvet_b\Cvet_l'\right)^{-1}\Cvet_l\Wvet_b & -\left(\Cvet_l\Wvet_b\Cvet_l'\right)^{-1}
\end{bmatrix},
\]
and after some algebra it is obtained the result:
\[
\widetilde{\bvet}^{(l)}_{\text{LCC}} = \widehat{\bvet} + \Wvet_b\Cvet_l'
\left(\Cvet_l\Wvet_b\Cvet_l'\right)^{-1}\left(\widehat{\avet}_l -\Cvet_l\widehat{\bvet}\right) , \quad l=1,\ldots,L.
\]
The previous expression can be re-stated as follows:
\[
\widetilde{\bvet}^{(l)}_{\text{LCC}} = \Lvet_l\widehat{\avet}_l + \Mvet_l\widehat{\bvet},
\]
with
$$
\Lvet_l = \Wvet_b\Cvet_l' \left(\Cvet_l\Wvet_b\Cvet_l'\right)^{-1}
\text{  e  }
\Mvet_l = \left[\Ivet_{n_b} - \Wvet_b\Cvet_l' \left(\Cvet_l\Wvet_b\Cvet_l'\right)^{-1}\Cvet_l\right] =
\left(\Ivet_{n_b} - \Lvet_l\Cvet_l\right) .
$$
It is thus:
\[
\begin{bmatrix}
	\widehat{\avet}_l \\[.25cm]
	\widetilde{\bvet}^{(l)}_{\text{LCC}}
\end{bmatrix} = \begin{bmatrix}
\Ivet_{n_l} & \Zerovet_{n_l\times n_b} \\[.25cm]
\Lvet_l & \Mvet_l
\end{bmatrix} \begin{bmatrix}
\widehat{\avet}_l \\[.25cm]
\widehat{\bvet}
\end{bmatrix} \quad \rightarrow \quad
\widetilde{\yvet}_{l,\text{LCC}} = \overline{\Mvet}_l \widehat{\yvet}_l ,
\quad
\text{with} \quad \overline{\Mvet}_l = \begin{bmatrix}
\Ivet_{n_l} & \Zerovet_{n_l\times n_b} \\[.25cm]
	\Lvet_l & \Mvet_l
\end{bmatrix}.
\]
Since, as can be easily checked, it is $\Mvet_l\Lvet_l = \Zerovet_{n_b \times n_b}$, the matrix $\overline{\Mvet}_l$ is idempotent ($\overline{\Mvet}_l\overline{\Mvet}_l=\overline{\Mvet}_l$). For, it is a projection matrix in a linear sub-space of ${\cal R}^{n_l+n_b}$ spanned by the relationship
$
\begin{bmatrix}
	\Ivet_{n_l} & \Zerovet_{n_l \times n_b} \\[.25cm]
	\Ivet_{n_l} & -\Cvet_l
\end{bmatrix}
\begin{bmatrix}
	\avet_l \\[.25cm] \bvet
\end{bmatrix} = \begin{bmatrix}
\widehat{\avet}_l \\[.25cm] \Zerovet_{n_b}
\end{bmatrix}
$.


\section*{Appendix 3. $LCC$ reconciliation with endogenous constraints for the toy example of Figure \ref{Fig:hts1}}
In this case the relationships linking the variable forming the hierarchy are:
\[
\begin{array}{rcl}
	T & = & X + Y = A + B + C + D + E \\
	X & = & A + B \\
	Y & = & C + D + E
\end{array} .
\]
Thus, there are two upper levels ($L=2$) for which it is possible to apply the $LCC$ reconciliation procedure with endogenous constraints:

\vspace{.5cm}

\noindent \underline{${l=1}$ ($n_1 = 1$, $n_b=5$)}
\begingroup\footnotesize
\[
\Wvet_1 = \begin{bmatrix}
	\sigma^2_T & 0 & 0 & 0 & 0 & 0 \\
	0 & \sigma^2_A & 0 & 0 & 0 & 0 \\
	0 & 0 & \sigma^2_B & 0 & 0 & 0 \\
	0 & 0 & 0 & \sigma^2_C & 0 & 0 \\
	0 & 0 & 0 & 0 &	\sigma^2_D & 0 \\
	0 & 0 & 0 & 0 & 0 & \sigma^2_E
\end{bmatrix}, \quad
\Uvet_1' = \begin{bmatrix}
	1 & -1 & -1 & -1 & -1 & -1
\end{bmatrix},
\]
\[
\Wvet_1\Uvet_1 =
\begin{bmatrix}
	\sigma^2_T \\ -\sigma^2_A \\ -\sigma^2_B \\ -\sigma^2_C \\ -\sigma^2_D \\ -\sigma^2_E
\end{bmatrix}, \quad
\Uvet'_1\Wvet_1\Uvet_1 = \sigma^2_T + \sigma^2_A + \sigma^2_B + \sigma^2_C +\sigma^2_D +\sigma^2_E .
\]
\endgroup
After a bit of algebra, it is found that the reconciled forecasts are given by:
\begingroup
\small
\[
\begin{array}{rcl}
	\widetilde{T}^{(1)}_{\text{en}} & = & \widehat{T} - \displaystyle\frac{\sigma^2_T}{\sigma^2_T+\sigma^2_A+\sigma^2_B+\sigma^2_C+\sigma^2_D+\sigma^2_E}
	\left(\widehat{T} - \widehat{A} - \widehat{B} - \widehat{C} - \widehat{D} - \widehat{E} \right) \\
	\widetilde{A}^{(1)}_{\text{en}} & = & \widehat{A} + \displaystyle\frac{\sigma^2_A}{\sigma^2_T+\sigma^2_A+\sigma^2_B+\sigma^2_C+\sigma^2_D+\sigma^2_E}
\left(\widehat{T} - \widehat{A} - \widehat{B} - \widehat{C} - \widehat{D} - \widehat{E} \right) \\
	\widetilde{B}^{(1)}_{\text{en}} & = & \widehat{B} + \displaystyle\frac{\sigma^2_B}{\sigma^2_T+\sigma^2_A+\sigma^2_B+\sigma^2_C+\sigma^2_D+\sigma^2_E}
\left(\widehat{T} - \widehat{A} - \widehat{B} - \widehat{C} - \widehat{D} - \widehat{E} \right) \\
	\widetilde{C}^{(1)}_{\text{en}} & = & \widehat{C} + \displaystyle\frac{\sigma^2_C}{\sigma^2_T+\sigma^2_A+\sigma^2_B+\sigma^2_C+\sigma^2_D+\sigma^2_E}
\left(\widehat{T} - \widehat{A} - \widehat{B} - \widehat{C} - \widehat{D} - \widehat{E} \right) \\
	\widetilde{D}^{(1)}_{\text{en}} & = & \widehat{D} + \displaystyle\frac{\sigma^2_D}{\sigma^2_T+\sigma^2_A+\sigma^2_B+\sigma^2_C+\sigma^2_D+\sigma^2_E}
\left(\widehat{T} - \widehat{A} - \widehat{B} - \widehat{C} - \widehat{D} - \widehat{E} \right) \\
	\widetilde{E}^{(1)}_{\text{en}} & = & \widehat{E} + \displaystyle\frac{\sigma^2_E}{\sigma^2_T+\sigma^2_A+\sigma^2_B+\sigma^2_C+\sigma^2_D+\sigma^2_E}
\left(\widehat{T} - \widehat{A} - \widehat{B} - \widehat{C} - \widehat{D} - \widehat{E} \right)	
\end{array} .
\]
\endgroup

\noindent Expressing each reconciled forecasts as a combination of the `direct' (base) forecast, and of the `implicit' forecast, obtained by applying the constraints to the base forecasts, gives:

\begingroup
\small
\[
\begin{array}{rcl}
	\widetilde{T}^{(1)}_{\text{en}} & = &
	\left(\displaystyle\frac{\sigma^2_A+\sigma^2_B+\sigma^2_C+\sigma^2_D+\sigma^2_E}{\sigma^2_T+\sigma^2_A+\sigma^2_B+\sigma^2_C+\sigma^2_D+\sigma^2_E}\right)\widehat{T} + 
	 \left(\displaystyle\frac{\sigma^2_T}{\sigma^2_T+\sigma^2_A+\sigma^2_B+\sigma^2_C+\sigma^2_D+\sigma^2_E}\right)\left(\widehat{A} + \widehat{B} + \widehat{C} + \widehat{D} + \widehat{E}\right) \\[.5cm]
	\widetilde{A}^{(1)}_{\text{en}} & = & 	\left(\displaystyle\frac{\sigma^2_T+\sigma^2_B+\sigma^2_C+\sigma^2_D+\sigma^2_E}{\sigma^2_T+\sigma^2_A+\sigma^2_B+\sigma^2_C+\sigma^2_D+\sigma^2_E}\right)\widehat{A} + 
	 \left(\displaystyle\frac{\sigma^2_A}{\sigma^2_T+\sigma^2_A+\sigma^2_B+\sigma^2_C+\sigma^2_D+\sigma^2_E}\right)\left(\widehat{T} - \widehat{B} - \widehat{C} - \widehat{D} - \widehat{E}\right) \\[.5cm]
\widetilde{B}^{(1)}_{\text{en}} & = & 	\left(\displaystyle\frac{\sigma^2_T+\sigma^2_A+\sigma^2_C+\sigma^2_D+\sigma^2_E}{\sigma^2_T+\sigma^2_A+\sigma^2_B+\sigma^2_C+\sigma^2_D+\sigma^2_E}\right)\widehat{B} + 
 \left(\displaystyle\frac{\sigma^2_B}{\sigma^2_T+\sigma^2_A+\sigma^2_B+\sigma^2_C+\sigma^2_D+\sigma^2_E}\right)\left(\widehat{T} - \widehat{A} - \widehat{C} - \widehat{D} - \widehat{E}\right) \\[.5cm]
\widetilde{C}^{(1)}_{\text{en}} & = & 	\left(\displaystyle\frac{\sigma^2_T+\sigma^2_A+\sigma^2_B+\sigma^2_D+\sigma^2_E}{\sigma^2_T+\sigma^2_A+\sigma^2_B+\sigma^2_C+\sigma^2_D+\sigma^2_E}\right)\widehat{C} + 
 \left(\displaystyle\frac{\sigma^2_C}{\sigma^2_T+\sigma^2_A+\sigma^2_B+\sigma^2_C+\sigma^2_D+\sigma^2_E}\right)\left(\widehat{T} - \widehat{A} - \widehat{B} - \widehat{D} - \widehat{E}\right) \\[.5cm]
\widetilde{D}^{(1)}_{\text{en}} & = & 	\left(\displaystyle\frac{\sigma^2_T+\sigma^2_A+\sigma^2_B+\sigma^2_C+\sigma^2_E}{\sigma^2_T+\sigma^2_A+\sigma^2_B+\sigma^2_C+\sigma^2_D+\sigma^2_E}\right)\widehat{D} + 
 \left(\displaystyle\frac{\sigma^2_D}{\sigma^2_T+\sigma^2_A+\sigma^2_B+\sigma^2_C+\sigma^2_D+\sigma^2_E}\right)\left(\widehat{T} - \widehat{A} - \widehat{B} - \widehat{C} - \widehat{E}\right) \\[.5cm]
\widetilde{E}^{(1)}_{\text{en}} & = & 	\left(\displaystyle\frac{\sigma^2_T+\sigma^2_A+\sigma^2_B+\sigma^2_C+\sigma^2_D}{\sigma^2_T+\sigma^2_A+\sigma^2_B+\sigma^2_C+\sigma^2_D+\sigma^2_E}\right)\widehat{E} + 
 \left(\displaystyle\frac{\sigma^2_E}{\sigma^2_T+\sigma^2_A+\sigma^2_B+\sigma^2_C+\sigma^2_D+\sigma^2_E}\right)\left(\widehat{T} - \widehat{A} - \widehat{B} - \widehat{C} - \widehat{D} \right)
\end{array}
\]

\vspace{.5cm}

\noindent\underline{${l=2}$ ($n_2 = 2$, $n_b=5$)}
\[
\Wvet_2 = \begin{bmatrix}
	\sigma^2_X & 0 & 0 & 0 & 0 & 0 & 0\\
	0 & \sigma^2_Y & 0 & 0 & 0 & 0 & 0 \\
	0 & 0 & \sigma^2_A & 0 & 0 & 0 & 0 \\
	0 & 0 & 0 & \sigma^2_B & 0 & 0 & 0 \\
	0 & 0 & 0 & 0 &	\sigma^2_C & 0 & 0 \\
	0 & 0 & 0 & 0 & 0 & \sigma^2_D & 0 \\
	0 & 0 & 0 & 0 & 0 & 0 & \sigma^2_E \\
\end{bmatrix}, \quad
\Uvet_2' = \begin{bmatrix}
	1 & 0 & -1 & -1 & 0 & 0 & 0 \\
	0 & 1 & 0 & 0 & -1 & -1 & -1
\end{bmatrix},
\]
\[
\Wvet_2\Uvet_2 =
\begin{bmatrix}
	\sigma^2_X  & 0 \\ 
	0 & \sigma^2_Y \\ 
	-\sigma^2_A & 0 \\ 
	-\sigma^2_B & 0 \\ 
	0 & -\sigma^2_C \\ 
	0 & -\sigma^2_D \\ 
	0 & -\sigma^2_E
\end{bmatrix}, \quad
\Uvet'_2\Wvet_2\Uvet_2 = \begin{bmatrix}
	\sigma^2_X + \sigma^2_A + \sigma^2_B & 0 \\
	0 & \sigma^2_Y  + \sigma^2_C +\sigma^2_D +\sigma^2_E
\end{bmatrix} .
\]
\endgroup

\noindent After a bit of algebra, we find that the reconciled forecasts are equal to
\begingroup
\small
\[
\begin{array}{rcl}
\widetilde{X}^{(2)}_{\text{en}} & = & \widehat{X} - \displaystyle\frac{\sigma^2_X}{\sigma^2_X + \sigma^2_A + \sigma^2_B}
\left(\widehat{X} - \widehat{A} - \widehat{B} \right) \\[.5cm]
\widetilde{Y}^{(2)}_{\text{en}} & = & \widehat{Y} - \displaystyle\frac{\sigma^2_Y}{\sigma^2_Y + \sigma^2_C + \sigma^2_D + \sigma^2_E}
\left(\widehat{Y} - \widehat{C} - \widehat{D} - \widehat{E} \right) \\[.5cm]
\widetilde{A}^{(2)}_{\text{en}} & = & \widehat{A} + \displaystyle\frac{\sigma^2_X}{\sigma^2_X + \sigma^2_A + \sigma^2_B}
\left(\widehat{X} - \widehat{B} \right) \\[.5cm]
\widetilde{B}^{(2)}_{\text{en}} & = & \widehat{B} + \displaystyle\frac{\sigma^2_X}{\sigma^2_X + \sigma^2_A + \sigma^2_B}
\left(\widehat{X} - \widehat{A} \right) \\[.5cm]
\widetilde{C}^{(2)}_{\text{en}} & = & \widehat{C} + \displaystyle\frac{\sigma^2_Y}{\sigma^2_Y + \sigma^2_C  + \sigma^2_D + \sigma^2_E}
\left(\widehat{Y} - \widehat{D} - \widehat{E} \right) \\[.5cm]
\widetilde{D}^{(2)}_{\text{en}} & = & \widehat{D} + \displaystyle\frac{\sigma^2_Y}{\sigma^2_Y + \sigma^2_C  + \sigma^2_D + \sigma^2_E}
\left(\widehat{Y} - \widehat{C} - \widehat{E} \right) \\[.5cm]
\widetilde{E}^{(2)}_{\text{en}} & = & \widehat{E} + \displaystyle\frac{\sigma^2_Y}{\sigma^2_Y + \sigma^2_C  + \sigma^2_D + \sigma^2_E}
\left(\widehat{Y} - \widehat{C} - \widehat{D} \right)
\end{array} 
\]
\endgroup
or equivalently:
\begingroup
\small
\[
\begin{array}{rcrcl}
\widetilde{X}^{(2)}_{\text{en}} & = &
\left(\displaystyle\frac{\sigma^2_A + \sigma^2_B}{\sigma^2_X + \sigma^2_A + \sigma^2_B}\right)\widehat{X} &+&
\left(\displaystyle\frac{\sigma^2_X}{\sigma^2_X + \sigma^2_A + \sigma^2_B}\right)\left(\widehat{A} + \widehat{B}\right) \\[.5cm]
\widetilde{Y}^{(2)}_{\text{en}} & = &
\left(\displaystyle\frac{\sigma^2_C + \sigma^2_D + \sigma^2_E}{\sigma^2_Y + \sigma^2_C + \sigma^2_D + \sigma^2_E}\right)\widehat{Y} &+&
\left(\displaystyle\frac{\sigma^2_Y}{\sigma^2_Y + \sigma^2_C + \sigma^2_D + \sigma^2_E}\right)\left(\widehat{C} + \widehat{D} + \widehat{E}\right) \\[.5cm]
\widetilde{A}^{(2)}_{\text{en}} & = &
\left(\displaystyle\frac{\sigma^2_X +\sigma^2_B}{\sigma^2_X + \sigma^2_A + \sigma^2_B}\right)\widehat{A} &+&
\left(\displaystyle\frac{\sigma^2_A}{\sigma^2_X + \sigma^2_A + \sigma^2_B}\right)\left(\widehat{X} - \widehat{B}\right) \\[.5cm]
\widetilde{B}^{(2)}_{\text{en}} & = &
\left(\displaystyle\frac{\sigma^2_X +\sigma^2_A}{\sigma^2_X + \sigma^2_A + \sigma^2_B}\right)\widehat{B} &+&
\left(\displaystyle\frac{\sigma^2_B}{\sigma^2_X + \sigma^2_A + \sigma^2_B}\right)\left(\widehat{X} - \widehat{A}\right) \\[.5cm]
\widetilde{C}^{(2)}_{\text{en}} & = &
\left(\displaystyle\frac{\sigma^2_Y + \sigma^2_D + \sigma^2_E}{\sigma^2_Y + \sigma^2_C + \sigma^2_D + \sigma^2_E}\right)\widehat{C} &+&
\left(\displaystyle\frac{\sigma^2_C}{\sigma^2_Y + \sigma^2_C + \sigma^2_D + \sigma^2_E}\right)\left(\widehat{Y} - \widehat{D} - \widehat{E}\right) \\[.5cm]
\widetilde{D}^{(2)}_{\text{en}} & = &
\left(\displaystyle\frac{\sigma^2_Y + \sigma^2_C + \sigma^2_E}{\sigma^2_Y + \sigma^2_C + \sigma^2_D + \sigma^2_E}\right)\widehat{D} &+&
\left(\displaystyle\frac{\sigma^2_D}{\sigma^2_Y + \sigma^2_C + \sigma^2_D + \sigma^2_E}\right)\left(\widehat{Y} - \widehat{C} - \widehat{E}\right) \\[.5cm]
\widetilde{E}^{(2)}_{\text{en}} & = &
\left(\displaystyle\frac{\sigma^2_Y + \sigma^2_C + \sigma^2_D}{\sigma^2_Y + \sigma^2_C + \sigma^2_D + \sigma^2_E}\right)\widehat{E} &+&
\left(\displaystyle\frac{\sigma^2_E}{\sigma^2_Y + \sigma^2_C + \sigma^2_D + \sigma^2_E}\right)\left(\widehat{Y} - \widehat{C} - \widehat{D}\right) \\
\end{array} .
\]
\endgroup

\vspace{-.5cm}

\bibliography{mybibfile}

\vspace{-.2cm}
\section*{Acknowldgements}
\vspace{-.3cm}
\noindent \textit{We would like to thank Fotios Petropoulos for his useful feedback on an earlier version of the manuscript.}

\end{document}